\documentclass[12pt,a4paper]{article}
\pdfoutput=1
\usepackage[format=hang]{caption}
\usepackage{epsfig,amssymb,amsmath,graphicx,subcaption,verbatim,hyperref,xcolor,ulem,epstopdf,psfrag,pstool,braket,array,enumerate,multirow,wrapfig}
\usepackage{graphics,color}

\numberwithin{equation}{section}

\begin{document} 

\newcommand{\be}{\begin{equation}}
\newcommand{\ee}{\end{equation}}
\newcommand{\bea}{\begin{eqnarray}}
\newcommand{\eea}{\end{eqnarray}}
\newcommand{\bean}{\begin{eqnarray*}}
\newcommand{\eean}{\end{eqnarray*}}
\font\upright=cmu10 scaled\magstep1
\font\sans=cmss12
\newcommand{\ssf}{\sans}
\newcommand{\stroke}{\vrule height8pt width0.4pt depth-0.1pt}
\newcommand{\Z}{\hbox{\upright\rlap{\ssf Z}\kern 2.7pt {\ssf Z}}}
\newcommand{\ZZ}{\Z\hskip -10pt \Z_2}
\newcommand{\C}{{\rlap{\upright\rlap{C}\kern 3.8pt\stroke}\phantom{C}}}
\newcommand{\R}{\hbox{\upright\rlap{I}\kern 1.7pt R}}
\newcommand{\HH}{\hbox{\upright\rlap{I}\kern 1.7pt H}}
\newcommand{\CP}{\hbox{\C{\upright\rlap{I}\kern 1.5pt P}}}
\newcommand{\identity}{{\upright\rlap{1}\kern 2.0pt 1}}
\newcommand{\half}{\frac{1}{2}}
\newcommand{\quart}{\frac{1}{4}}
\newcommand{\pr}{\partial}
\newcommand{\bm}{\boldmath}
\newcommand{\I}{{\cal I}} 
\newcommand{\M}{{\cal M}}
\newcommand{\N}{{\cal N}}
\newcommand{\e}{\varepsilon}
\newcommand{\ep}{\epsilon}
\newcommand{\balpha}{\mbox{\boldmath $\alpha$}}
\newcommand{\bgamma}{\mbox{\boldmath $\gamma$}}
\newcommand{\blambda}{\mbox{\boldmath $\lambda$}}
\newcommand{\bep}{\mbox{\boldmath $\varepsilon$}}
\newcommand{\Oh}{{\rm O}}
\newcommand{\x}{{\bf x}}
\newcommand{\y}{{\bf y}}
\newcommand{\X}{{\bf X}}
\newcommand{\Y}{{\bf Y}}
\newcommand{\z}{{\bar z}}
\newcommand{\w}{{\bar w}}
\newcommand{\tT}{{\tilde T}}
\newcommand{\tX}{{\tilde\X}}
\def\ir3{\int_{\mathbb{R}^{3}}}

\thispagestyle{empty}
\rightline{DAMTP-2017-43}
\vskip 3em
\begin{center}
{{\bf \Large Quantized Skyrmions from SU(4) Weight Diagrams
}} 
\\[15mm]

{\bf \large C.~J. Halcrow\footnote{email: christ@impcas.ac.cn}} \\[1pt]
\vskip 1em
{\it 
Institute of Modern Physics, \\ 
Chinese Academy of Sciences,\\
 Lanzhou 730000, China,} \\[20pt]

{\bf \large N.~S. Manton\footnote{email: N.S.Manton@damtp.cam.ac.uk}
  and J.~I. Rawlinson\footnote{email: jir25@damtp.cam.ac.uk}} \\[1pt]
\vskip 1em
{\it 
Department of Applied Mathematics and Theoretical Physics,\\
University of Cambridge, \\
Wilberforce Road, Cambridge CB3 0WA, U.K.}
\vspace{12mm}

\abstract{}

Starting from solutions of the lightly-bound Skyrme model, we construct many new Skyrmion solutions of the standard Skyrme model with tetrahedral or octahedral symmetry. These solutions are closely related to weight diagrams of the group SU(4), which enables us to systematically derive some geometric and energetic properties of the Skyrmions, up to baryon number 85. We discuss the rigid body quantization of these Skyrmions, and compare the results with properties of a selection of observed nuclei.   

\end{center}

\vskip 5pt

\vfill
\newpage
\setcounter{page}{1}
\renewcommand{\thefootnote}{\arabic{footnote}}


\section{Introduction} 

The lightly-bound Skyrme model, developed by the Leeds group \cite{Har, GHS, GHKMS}, gives new insight into the structure and symmetries of Skyrmions for a large range of baryon numbers, $B$. In this model, Skyrmion solutions are very well approximated by clusters of $B=1$ Skyrmions located at the points of a face centred cubic (FCC) lattice. Four different Skyrmion orientations occur periodically in the lattice. The Skyrmions do not significantly merge in the lightly-bound model, and the classical binding energy of each Skyrmion is small compared to its rest mass. This is an attractive feature, analogous to the small binding energy of nucleons in nuclei, of order 8 MeV per nucleon, compared to the nucleon rest mass 938 MeV. However this feature is considerably spoiled by quantization, where the spin energy of each $B=1$ Skyrmion adds significantly to the total energy. Also, the Lagrangian of the lightly-bound model is quite complicated. 

We do not work directly with the lightly-bound model, but instead work with the standard Skyrme model, with its sigma model term, Skyrme term, and pion mass term \cite{Sky,AN}. The Skyrme field is an SU(2)-valued field
\be
U({\bf{x}}) = \sigma({\bf{x}}) \, {\bf 1} 
+ i\boldsymbol{\pi}({\bf{x}}) \cdot \boldsymbol{\tau} \,,
\ee
where ${\bf 1}$ is the unit matrix and $\boldsymbol{\tau}$ are Pauli matrices. $\sigma({\bf{x}})$ and $\boldsymbol{\pi}({\bf{x}})$ are sigma and pion fields satisfying $\sigma^2 + \boldsymbol{\pi} \cdot  \boldsymbol{\pi} = 1$. The classical vacuum configuration, and also the boundary condition at spatial infinity for Skyrmions, is $U = {\bf 1}$.

The baryon number and (static) energy of Skyrme field configurations are integrals involving the ``current'' $R_i = (\partial_i U)U^{-1}$. The baryon number $B$ is the topological degree of $U$,
\be
B = -\frac{1}{24\pi^2}\ir3 \varepsilon_{ijk}{\rm Tr}(R_iR_jR_k) \,
d^3x \,,
\ee
an integer that we assume to be positive. The energy (in Skyrme units, and conveniently normalised) is
\be
E = \frac{1}{12\pi^2}\ir3\bigg\{-{\frac{1}{2}}{\rm Tr}(R_iR_i) 
-{\frac{1}{16}}{\rm Tr}([R_i,R_j][R_i,R_j])
+ \, m^2{\rm Tr}({\bf 1} - U) \bigg\} d^3x \, ,
\ee
where $m$ is the dimensionless pion mass. We have set $m=1$ for the numerical calculations in this paper. Skyrmions are absolute minimisers of $E$ for each baryon number $B$, or in a looser sense, local minima of $E$ with energy close to the absolute minimum. 

We have found previously \cite{LM, HKM} that after appropriate calibration, quantized standard Skyrmions have reasonable spectra, matching those of various nuclei, including Carbon-12 and Oxygen-16. However, the clusters in the lightly-bound model are still very helpful as starting configurations to relax to the true Skyrmions, and we have observed that the true Skyrmions have largely unchanged shapes and symmetries. The main difference is that the $B=1$ Skyrmions merge slightly as they bind together, and the gaps between them acquire a small baryon and energy density. Several examples are illustrated in this paper.

In \cite{GHKMS} the optimal clusters, with strongest binding, were found for all baryon numbers up to $B=23$, and there is in principle no problem going to higher $B$. An exploration of larger clusters was initiated in \cite{Man}, where relationships to magic nuclei and the shell model were proposed. That work made clear that an interesting class of clusters are those corresponding to the 3-dimensional  weight diagrams of irreducible representations (irreps) of the Lie group SU(4). This observation builds on the insight of Wigner into the role of SU(4) symmetry and its irreps in nuclear physics \cite{Wig}. Wigner's ideas were developed by Cook, Dallacasa and others \cite{Coo, CD, Eve, Lez}, who viewed certain weight diagrams of SU(4) as illustrating the spatial structure of selected nuclei, especially those with baryon numbers $B=4,16,40,80,140$. This idea was rather speculative, but has found some justification in the context of Skyrmions \cite{Man}. An important detail is that if nucleons are placed at the locations of weights in a weight diagram, then the total number of nucleons $B$ is the number of weights counted {\sl without multiplicity}, i.e. the number of distinct weights. This differs from the dimension of the irrep, which counts weights {\sl with multiplicity}.
 
The mathematical theory of weight multiplicities is well developed \cite{Hum}, and it is known how to find the number of distinct weights of an SU(4) irrep \cite{Kas}. In Section 2, we present the formula for the number of  distinct weights $c$ in terms of the Dynkin indices $[P,Q,R]$ of the irrep. It is a rather more complicated polynomial than the formula for the dimension $d$ of the irrep. However, $c$ can be expressed as a simple combination of dimensions of irreps with Dynkin indices close to those for the irrep of interest. This is based on the idea that the weight multiplicities of an irrep can be reduced to unity by subtracting and adding weights of neighbouring irreps. The result generalises a formula for SU(3) irreps, where $c$ is simply the difference of two dimensions. 

\begin{figure}
	\centering
	\includegraphics[width=0.7\textwidth]{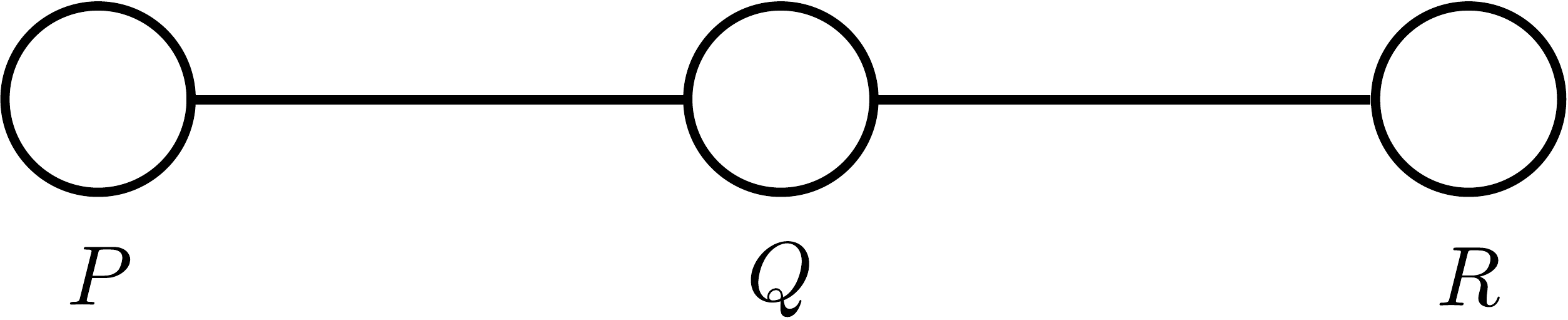} 
	\caption{The Dynkin diagram for SU(4).} \label{Dynkin}
\end{figure}

The weight lattice of SU(4) is a 3-dimensional body centred cubic (BCC) lattice, and weight diagrams are finite clusters of points of this lattice. All such clusters are symmetric under the Weyl group, which is the tetrahedral group T, and some clusters are symmetric under the octahedral (cubic) group O. The extra symmetry arises from the outer automorphism that exchanges the ends of the SU(4) Dynkin diagram. Clusters have symmetry O if the Dynkin indices are unchanged by this reflection. The Dynkin diagram, with Dynkin indices attached, is shown in Figure \ref{Dynkin}. A further property of weight clusters is that they are convex polyhedra and have no interior holes. Their faces are generically hexagons with alternating side lengths, and rectangles.

Weights of an irrep can only differ by elements of the root lattice, and the root lattice is an FCC sublattice of the weight lattice, of order four. Weights of an irrep are therefore characterised by a congruence class called the quadrality (generalising SU(3) triality), and lie in the coset of the FCC root lattice corresponding to the quadrality. So all weight clusters (possibly after a shift) are clusters of a fixed FCC lattice, as in the lightly-bound Skyrme model. Congruence class 0 coincides with the root lattice, and contains, for example, the weight diagram of the adjoint representation ${\bf 15}$. This diagram has 13 distinct weights, giving a model for an octahedrally symmetric $B=13$ Skyrmion. Congruence classes 1 and 3 have equivalent weight clusters related by the action of the outer automorphism; the fundamental irreps ${\bf 4}$ and ${\overline{\bf 4}}$ are in these classes, for example, and give isomorphic tetrahedral clusters modelling $B=4$ Skyrmions. Physically, these clusters differ in their orientation in space, but when we quantize Skyrmions we must allow for all orientations, so there is no distinction between congruence classes 1 and 3. Congruence class 2 contains further weight clusters. The smallest is the octahedral cluster of the irrep ${\bf 6}$, whose weights have unit multiplicity, so this cluster models a $B=6$ Skyrmion.

In Section 2 we tabulate systematically the weight clusters for all 
irreps up to baryon (cluster) number 85. These have baryon numbers
\be
B=1,4,6,10,13,16,19,20,28,31,35,38,40,44,50,52,55,56,68,79,80,84,85 \,,
\ee 
and all have tetrahedral symmetry at least. This is a more substantial list of examples than were discussed in \cite{GHKMS}, although it omits some interesting clusters that do not occur as weight diagrams. In Section 3 we discuss how a weight cluster becomes a Skyrmion when $B=1$ Skyrmions are located at the weight points, and their orientations are appropriately fixed. The cluster needs to be numerically relaxed, to obtain a Skyrmion of minimal energy. We present the numerically generated Skyrmions in Figures \ref{Q0sk} -- \ref{Q2sk}. It was pointed out in \cite{GHKMS} that some of the small clusters do not give particularly stable Skyrmions; however, for larger baryon numbers the clusters have more bonds between nearest neighbours, and they seem to be more stable, especially if they have truncated shapes rather than being pure tetrahedra or pure octahedra. 

Classical Skyrmions should be thought of as intrinsic structures of nuclei. They spontaneously break most spatial and internal symmetries, because they have a definite position and orientation, and pion field orientation, but these symmetries are restored by quantization. In Section 4 we consider the rigid body quantization of all these Skyrmions. The resulting quantum states are characterised by their spin, isospin and parity (and also momentum). The spin/isospin combinations are constrained by the residual unbroken symmetry of the classical Skyrmion, and also by the associated Finkelstein--Rubinstein (FR) sign representation that arises topologically. The general approach to rigid body quantization of Skyrmions is well established \cite{ANW,MMW}, but the details of the symmetry operations and FR signs are sometimes different here than in previous discussions. The tetrahedral (or octahedral) symmetry is realised in different ways depending on the congruence class. We determine the allowed low-lying states for each congruence class, finding their spins, isospins and parities. The calculations for congruence class 1 (class 3 is the same, so we generally don't mention it in what follows) are the same as those presented in \cite{Man}. Here the baryon number is always a multiple of 4, so the spin and isospin are integers. The analysis for classes 0 and 2 is not too hard, but the results are mostly novel. Clusters in class 0 have an odd baryon number, because there is always a single weight at the origin, and other weights lie in orbits of the tetrahedral group with an even number of weights. The quantized states therefore have half-integer spin and isospin. For congruence class 2, the baryon number is even but not a multiple of four. For congruence classes 0 and 2, the FR sign representation can be either trivial or non-trivial if there is octahedral symmetry. So there are various cases to consider. The allowed spin/isospin/parity states for all isospins up to $I=3$ are classified in detail in Section 5.

In Section 6 we determine the energy levels of the various states obtained from the rigid body quantization. This uses the information about the moments of inertia in space and isospace of the Skyrmions; because of the high degree of symmetry, each moment of inertia tensor has few independent entries. We also compare the predictions for the quantized Skyrmions with known nuclear states that have been established as lying in isospin multiplets. Though not all nuclear states can be explained convincingly, there are some surprising successes. In particular, certain states of the $B=31$ and $B=38$ Skyrmions match data rather well. For these baryon numbers, neither the Skyrmions nor their quantization have been considered before.

Section 7 contains our conclusions.

\section{SU(4) Weight Clusters and the Cluster Number}

The root lattice of SU(4) is an FCC lattice. We fix the scaling of this, so that the Cartesian coordinates $\x = (x,y,z)$ of the lattice points are even integers, satisfying $x+y+z = 0$ mod 4. The roots themselves are the 12 vectors $(\pm 2, \pm 2, 0)$, $(\pm 2, 0, \pm 2)$ and $(0, \pm 2, \pm 2)$. A basis of simple roots $\balpha_j : j=1,2,3$ is 
\be
\balpha_1 = (2,2,0), \, \quad \balpha_2 = (-2,0,2) \,, \quad \balpha_3 = 
(2,-2,0) \,.
\ee
As expected from the SU(4) Dynkin diagram, the angle between $\balpha_2$ and both $\balpha_1$ and $\balpha_3$ is $\frac{2\pi}{3}$, whereas $\balpha_1$ and $\balpha_3$ are orthogonal.

The fundamental weights $\blambda_i : i=1,2,3$ satisfy
\be
\blambda_i \cdot \balpha_j = 4 \delta_{ij} \,.
\ee
Explicitly, we find that
\be
\blambda_1 = (1,1,1) \,, \quad \blambda_2 = (0,0,2) \,, \quad
\blambda_3 = (1,-1,1) \,.
\ee
The simple roots can be expressed in terms of these as
\be
\balpha_1 = 2\blambda_1 - \blambda_2 \,, \quad 
\balpha_2 = 2\blambda_2 - \blambda_1 - \blambda_3 \,, \quad 
\balpha_3 = 2\blambda_3 - \blambda_2 \,.
\ee

The weight lattice is the integer span of the fundamental weights, so the general weight is $p\blambda_1 + q\blambda_2 + r\blambda_3$, with $p,q$ and $r$ integers. It is a BCC lattice, dual to the FCC root lattice. There are four cosets of the root lattice in the weight lattice, and they are labelled by a quadrality $0,1,2$ or $3$. The quadrality of a weight is $p + 2q + 3r$ mod 4. Note that the roots all have quadrality $0$. Weights of an irrep differ by elements of the root lattice, so they all have the same quadrality.

Each irrep has a highest weight $P\blambda_1 + Q\blambda_2 + R\blambda_3$, and $[P,Q,R]$ are known as the Dynkin indices of the irrep. Other weights of the irrep are obtained by subtracting positive combinations of $\balpha_1 \,, \balpha_2$ and $\balpha_3$, which doesn't affect the quadrality, so the quadrality of the irrep is $P + 2Q + 3R$ mod 4.

The smallest examples of irreps with quadralities $0,1,2,3$, respectively, are the trivial irrep ${\bf 1} = [0,0,0]$, the fundamental irrep ${\bf 4} = [1,0,0]$, the irrep ${\bf 6} = [0,1,0]$, and the antifundamental irrep $\overline{\bf 4} = [0,0,1]$. The notation shows the dimension (in bold) and the Dynkin indices. The $\bf 6$ is a tensor representation of SU(4), but it is also the fundamental vector representation of SO(6), whose Lie algebra is isomorphic to that of SU(4). 

The Dynkin diagram of SU(4) has a reflection symmetry, exchanging the ends. This corresponds to an outer automorphism of the algebra, namely, complex conjugation. It follows that the irreps with Dynkin indices $[P,Q,R]$ and $[R,Q,P]$ are conjugate and have the same dimension. Their weight diagrams are related by the reflection that exchanges $\blambda_1$ and $\blambda_3$. Since weight diagrams have further reflection symmetries, these conjugate weight diagrams just differ by a rotation. For example, the weight diagrams of the $\bf 4$ and $\overline{\bf 4}$ are differently oriented regular tetrahedra. As we interpret weight diagrams as Skyrmion clusters, all orientations have equal importance, so we can identify the weight diagrams of $[P,Q,R]$ and $[R,Q,P]$. The irreps with $P=R$ are self-conjugate.

The weight diagram of any SU(4) irrep is invariant under the Weyl group, the group generated by reflections in the planes orthogonal to the simple roots. For SU(4), the Weyl group is the tetrahedral group T. This is the full symmetry group of the tetrahedron, with 24 elements, and is a subgroup of O(3). Abstractly it is the permutation group $S_4$, which permutes the vertices of a regular tetrahedron. For the self-conjugate irreps there is a further symmetry, a rotation by $\frac{\pi}{2}$, and the weight diagrams of these irreps are invariant under O, the full octahedral subgroup of O(3), with 48 elements.

These symmetries are important for us -- they are the body-fixed symmetry groups of the Skyrmion clusters, and have a crucial influence on the allowed spin/isospin/parity of quantum states.

Skyrmions are free to move, so the position of the centre of a Skyrmion does not have to coincide with the origin of Cartesian coordinates. Weight diagrams of quadrality $0$ are subsets of the FCC root lattice. On the other hand, weight diagrams of non-zero quadrality are subsets of shifted FCC root lattices. For example, weights of the quadrality $1$ irrep ${\bf 4}$ form a regular tetrahedron with centre of mass at the origin, but only after a shift do they form a cluster in the FCC root lattice with one point at the origin, for example, as the set of points $(0,0,0)$, $(2,2,0)$, $(2,0,2)$ and $(0,2,2)$. The first description, as a weight cluster, is more convenient here, where we use some representation theory, and quadrality is significant, but the second description is how the cluster appears in \cite{GHKMS}. 

For us, the most important characteristic of a weight diagram is the number of distinct weights it has. We call this the cluster number $c$ of the irrep, and it is the same as the baryon number $B$ of the corresponding Skyrmion. For the weight diagram with Dynkin indices $[P,Q,R]$, the cluster number is \cite{Kas}
\bea
c(P,Q,R) &=& \frac{2}{3}Q^3 + 2Q^2 + \frac{7}{3}Q + 1   
+ \left(2Q^2 + 4Q + \frac{11}{6} \right)(P+R) \nonumber \\ 
&& + (Q + 1)(P+R)^2 + (4Q + 1)PR + \frac{1}{6}(P+R)^3 + (P+R)PR \,.
\label{clustSU4}
\eea
This formula can be verified by a truncation argument. Any weight diagram can be obtained by starting with a complete, pure tetrahedral weight diagram, then truncating four equal, smaller tetrahedra from each vertex, and finally truncating the six remaining edges symmetrically. To get the weight diagram of $[P,Q,R]$, with $P \ge R$, one needs to start with the pure tetrahedral diagram of the irrep with Dynkin indices $[P+2Q+3R,0,0]$, which has $P+2Q+3R+1$ weights along an edge (and edge length $P+2Q+3R$). The first truncation removes tetrahedra with $Q+2R$ weights along an edge, and the second truncation removes $R$ rectangular layers. The resulting diagram has rectangular faces with edge lengths $P$ and $R$, hexagonal faces with alternating edge lengths $P$ and $Q$, and hexagonal faces with alternating edge lengths $Q$ and $R$. Some faces degenerate to lines, triangles or points if one or two of $P$, $Q$ and $R$ are zero. 

The polynomial $c(P,Q,R)$ is cubic and rather complicated. It can be expressed more simply as a linear combination of the dimension of the irrep $d(P,Q,R)$ and the dimensions of a few irreps with nearby Dynkin indices \cite{Kas}.

Let us recall how this works for SU(3). The SU(3) irrep with Dynkin indices $[P,Q]$ has a hexagonal weight diagram, with side lengths $P$ and $Q$. This diagram is obtained from an equilateral triangle of side length $P+2Q$ by truncating three triangles of side length $Q-1$. As the $n$th triangular number is $\frac{1}{2}n(n+1)$, the cluster number, the number of distinct weights, is
\be
c(P,Q) = \frac{1}{2}(P^2 + Q^2 + 4PQ + 3P + 3Q + 2) \,,
\label{clustSU3}
\ee
symmetric under exchange of $P$ and $Q$. 

The diagram has weights lying on a nested set of hexagons, with the inner hexagons degenerating to triangles if $P \ne Q$. The weight multiplicity is $1$ on the outside and increases by $1$ at each step inwards until one reaches a triangle, after which it remains constant. The cluster number $c(P,Q)$ is therefore the difference between the dimensions of the irreps with Dynkin indices $[P,Q]$ and
$[P-1,Q-1]$,
\be
c(P,Q) = d(P,Q) - d(P-1,Q-1) \,,
\label{clustcombSU3}
\ee
because the distinct weights of the irrep with Dynkin indices $[P-1,Q-1]$ coincide with those for the irrep with Dynkin indices $[P,Q]$, except for their absence on the outer hexagon, and the multiplicities are all less by 1. Note that the shift vector from the highest weight of the irrep $[P,Q]$ to the highest weight of the irrep $[P-1,Q-1]$ is the negative of the positive root $\balpha_1 + \balpha_2$, where $\balpha_1$ and $\balpha_2$ are SU(3) simple roots.

Using the formula \cite{McP}
\be
d(P,Q) = \frac{1}{2}(P+1)(Q+1)(P+Q+2)
\label{dimSU3}
\ee
for the dimension of an SU(3) irrep in (\ref{clustcombSU3}), we easily verify (\ref{clustSU3}). Note that $c(P,Q)$ and $d(P,Q)$ are na\"{\i}vely only defined for non-negative integers $P$ and $Q$. But formula (\ref{clustcombSU3}) is true even if $P$ or $Q$ is zero, making $P-1$ or $Q-1$ equal to $-1$. In either case $d(P-1,Q-1) = 0$, and $c(P,Q) = d(P,Q)$. This is correct because if $P$ or $Q$ is zero, the weight diagram is triangular and all weights have multiplicity 1. Note also that the irreps whose dimensions are combined in (\ref{clustcombSU3}) have the same triality, as $P+2Q = (P-1) + 2(Q-1)$ mod 3.

For an SU(4) irrep the dimension is \cite{McP}
\be
d(P,Q,R) = \frac{1}{12}(P+1)(Q+1)(R+1)(P+Q+2)(Q+R+2)(P+Q+R+3) \,,
\label{dimSU4}
\ee
a polynomial of degree 6 in $(P,Q,R)$, symmetric under exchange of $P$ and $R$. The dimension of an irrep with Dynkin indices shifted from $[P,Q,R]$ by fixed amounts is another polynomial of degree 6. By combining these polynomials for a suitable, finite set of shifts we obtain the cubic polynomial $c(P,Q,R)$. The combination preserves the exchange symmetry, and each shift preserves the quadrality $P+2Q+3R$ mod 4.

The correct combination is \cite{Kas}
\bea
c(P,Q,R) &=& d(P,Q,R) \nonumber \\
&& - d(P-1,Q-1,R+1) - d(P+1,Q-1,R-1) - d(P-1,Q,R-1) \nonumber \\
&& + d(P,Q-2,R) + d(P-2,Q-1,R) + d(P,Q-1,R-2) \nonumber \\
&& - d(P-1,Q-2,R-1) \,.
\label{clustcombSU4}
\eea
The seven shift vectors are the negatives of the combinations of simple roots
\bea
&& \balpha_1 + \balpha_2 \,, \quad \balpha_2 + \balpha_3 \,,
\quad \balpha_1 + \balpha_2 + \balpha_3 \,, 
\quad \balpha_1 + 2\balpha_2 + \balpha_3 \,,
\nonumber \\
&& \ 2\balpha_1 + 2\balpha_2 + \balpha_3 \,,
\quad \balpha_1 + 2\balpha_2 + 2\balpha_3 \,,
\quad 2\balpha_1 + 3\balpha_2 + 2\balpha_3 \,. 
\eea
The first three of these are the non-simple positive roots $\bgamma_1$, $\bgamma_2$ and $\bgamma_3$, and the remaining combinations are their sums $\bgamma_1 + \bgamma_2$, $\bgamma_1 + \bgamma_3$, $\bgamma_2 + \bgamma_3$ and $\bgamma_1 + \bgamma_2 + \bgamma_3$. 

In the relation (\ref{clustcombSU4}) the coefficients are $1$ or $-1$, depending on whether the shift involves an even or odd number of the $\bgamma_i$. Some of these terms can fail to be true dimensions of irreps if $P,Q$ or $R$ have value 0, so that $P-2$, $Q-2$ or $R-2$ have value $-2$. The corresponding dimension $d$ may then be negative, but it still makes an essential contribution. If $P-2$, $Q-2$ or $R-2$ have value $-1$ then $d=0$.

An example is the cluster number for Dynkin indices $[0,Q,0]$. The weight diagram is a pure octahedron. The relation (\ref{clustcombSU4}) simplifies to
\be
c(0,Q,0) = d(0,Q,0) + d(0,Q-2,0) + d(-2,Q-1,0) + d(0,Q-1,-2) \,,
\ee
where
\bea
&& d(0,Q,0) = \frac{1}{12}(Q+1)(Q+2)^2(Q+3) \,, \quad
d(0,Q-2,0) = \frac{1}{12}(Q-1)Q^2(Q+1) \,, \nonumber \\
&& \quad\quad\quad d(-2,Q-1,0) = d(0,Q-1,-2) = -\frac{1}{12}(Q-1)Q^2(Q+1) \,.
\eea
Therefore
\be
c(0,Q,0) = \frac{1}{3}(Q + 1)(2Q^2 + 4Q + 3) \,,
\ee
with values $6,19,44$ and $85$ for $Q=1,2,3$ and $4$.

Another example is the truncated octahedron with Dynkin indices $[1,1,1]$. Here the cluster number is 38, whereas the dimension is 64. This Skyrmion cluster appears to be particularly stable for its size, because the vertex of the cluster located at the highest weight has six nearest neighbours, which are reached from the vertex by subtracting each of the six positive roots, whereas for other clusters (e.g. a pure tetrahedron) the highest weight vertex has fewer nearest neighbours. All $B=1$ Skyrmions in the $B=38$ cluster therefore have six or more nearest neighbours, making it energetically unfavourable to pull one away. The conical angle at the vertex is also larger (less pointed) than for other types of vertex, and in compensation there are more vertices, namely 24 of them, in order to satisfy the polyhedral version of the Gauss--Bonnet theorem.

Note that the clusters with quadrality 0 all have $c$ odd, whereas those of quadralities 1 and 2 have $c$ even. This is because almost all orbits of the tetrahedral Weyl group acting on the weight lattice have an even number of points (24, 12, 6, or 4), and any cluster is a disjoint union of orbits. The exceptional odd orbit is the point at the origin, which occurs in all weight diagrams of quadrality 0, but not in those of quadralities 1 or 2. Orbits with six points only occur in the clusters of quadralities 0 and 2. The clusters of quadrality 1 therefore have $c$ a multiple of 4, and at their core is the tetrahedron of the fundamental irrep. Consistent with this, none of the quadrality 1 clusters can have symmetry O. This is a consequence of $P$ and $R$ necessarily being different, because if $P=R$ then the quadrality is $2Q$ mod 4, which is even.

We conclude this section by tabulating the weight clusters of SU(4) irreps, up to cluster number 85. They are grouped by their quadrality. For quadralities $0$ and $2$ we assume that $P \ge R$. If $P > R$ there is a second cluster with $P < R$ and the same quadrality, but this is just a copy of the first rotated by $\frac{\pi}{2}$, and is not tabulated. For quadrality $1$ we do not restrict to $P \ge R$. In this case, exchanging $P$ and $R$ results in an equivalent cluster with quadrality 3, rotated by $\frac{\pi}{2}$.

The tables give the Dynkin indices $[P,Q,R]$, the cluster (baryon) number $c(P,Q,R)$ (clusters with the same cluster number are distinguished by subscripts including the quadrality), the dimension $d(P,Q,R)$, and the symmetry  of the cluster (either the tetrahedral group T or octahedral group O). When the symmetry group of a cluster is O there are two possibilities for the Finkelstein--Rubinstein signs. What these mean and how they are calculated are explained in Section 4. The two cases are denoted  ${\rm O}+$ and ${\rm O}-$. The last four columns give the number of nearest neighbour bonds (short bonds) $N_{\rm short}$, the number of next-to-nearest neighbour bonds (long bonds) $N_{\rm long}$, and a (dimensionless) estimate of the binding energy per baryon $E_{\rm bond}/B$, where
\be
E_{\rm bond} = N_{\rm short} - 0.8 \, N_{\rm long} \,.
\ee 

In the lightly-bound model, $N_{\rm short} - 0.5 \, N_{\rm long}$ gives a good estimate for the binding energy of a cluster \cite{Har2}. $E_{\rm bond}$ works better in the standard Skyrme model except for some exceptional Skyrmions that have particularly visible clustering into $B=4$ cubes and $B=3$ tetrahedra, notably the $B=4$ and $B=28$ Skyrmions. The binding energy is significantly greater in these cases.
 
\begin{table}
	\begin{center}
		\begin{tabular}{|c|c|c|c|c|c|c|}
			\hline
			$[P,Q,R]$ & $c(P,Q,R)$ & $d(P,Q,R)$ & Symmetry 
			& $N_{\rm short}$ & $N_{\rm long}$ & $E_{\rm bond}/B$ \\ \hline
			[0,0,0] & 1 & 1 & ${\rm O}+$ & 0 & 0 & 0.0 \\ \hline
			[1,0,1] & 13 & 15 & ${\rm O}-$ & 36 & 12 & 2.03 \\ \hline
			[0,2,0] & 19 & 20 & ${\rm O}+$ & 60 & 18 & 2.4 \\ \hline
			[2,1,0] & 31 & 45 & T & 108 & 30 & 2.71 \\ \hline
			[4,0,0] & 35 & 35 & T & 120 & 30 & 2.74 \\ \hline
			[2,0,2] & 55 & 84 & ${\rm O}-$ & 216 & 90 & 2.62 \\ \hline
			[1,2,1] & 79 & 175 & ${\rm O}-$ & 336 & 126 & 2.98 \\ \hline
			[0,4,0] & 85 & 105 & ${\rm O}+$ & 360 & 132 & 2.99 \\ \hline
		\end{tabular}
		\vskip 7pt
		\caption{Weight Clusters with Quadrality 0.}
		\label{Quad0table}
	\end{center}	
\end{table}

\begin{table}
	\begin{center}
		\begin{tabular}{|c|c|c|c|c|c|c|}
			\hline
			$[P,Q,R]$ & $c(P,Q,R)$ & $d(P,Q,R)$ & Symmetry & $N_{\rm short}$ 
			& $N_{\rm long}$ & $E_{\rm bond}/B$ \\ \hline
			[1,0,0] & 4 & 4 & T & 6 & 0 & 1.5 \\ \hline
			[0,1,1] & 16 & 20 & T & 48 & 12 & 2.4 \\ \hline
			[0,0,3] & 20 & 20 & T & 60 & 12 & 2.52 \\ \hline
			[2,0,1] & 28 & 36 & T & 96 & 36 & 2.4 \\ \hline
			[1,2,0] & 40 & 60 & T & 150 & 48 & 2.79 \\ \hline
			[3,1,0] & 52 & 84 & T & 198 & 60 & 2.88 \\ \hline
			[5,0,0] & 56 & 56 & T & 210 & 60 & 2.89 \\ \hline
			[1,1,2] & $68_1$ & 140 & T & 282 & 108 & 2.88 \\ \hline
			[1,0,4] & $80_{1a}$ & 120 & T & 330 & 132 & 2.81 \\ \hline
			[0,3,1] & $80_{1b}$ & 140 & T & 336 & 120 & 3.0 \\ \hline
		\end{tabular}
		\vskip 7pt
		\caption{Weight Clusters with Quadrality 1.}
		\label{Quad1table}
	\end{center}	
\end{table}

\begin{table}
	\begin{center}
		\begin{tabular}{|c|c|c|c|c|c|c|}
			\hline
			$[P,Q,R]$ & $c(P,Q,R)$ & $d(P,Q,R)$ & Symmetry & $N_{\rm short}$ 
			& $N_{\rm long}$ & $E_{\rm bond}/B$ \\ \hline
			[0,1,0] & 6 & 6 & ${\rm O}-$ & 12 & 3 & 1.6 \\ \hline
			[2,0,0] & 10 & 10 & T & 24 & 3 & 2.16 \\ \hline
			[1,1,1] & 38 & 64 & ${\rm O}-$ & 144 & 51 & 2.72 \\ \hline
			[0,3,0] & 44 & 50 & ${\rm O}+$ & 168 & 57 & 2.78 \\ \hline
			[3,0,1] & 50 & 70 & T & 192 & 75 & 2.64 \\ \hline
			[2,2,0] & $68_2$ & 126 & T & 276 & 93 & 2.96 \\ \hline
			[4,1,0] & $80_2$ & 140 & T & 324 & 105 & 3.0 \\ \hline
			[6,0,0] & 84 & 84 & T & 336 & 105 & 3.0 \\ \hline
		\end{tabular}
		\vskip 7pt
		\caption{Weight Clusters with Quadrality 2.}
		\label{Quad2table}
	\end{center}
\end{table}

The formula for the number of nearest neighbour (short) bonds is
\bea
N_{\rm short}(P,Q,R) &=& 4Q^3+6Q^2+2Q + (12Q^2+12Q+2)(P+R) 
+ (6Q+3)(P+R)^2 \nonumber \\
&& \ + 24QPR + (P+R)^3 + 6(P+R)PR \,.
\label{shortbonds}
\eea
We found this by a truncation argument. In the infinite FCC lattice, each point has 12 nearest neighbours, so there are 6 short bonds per baryon. In a finite cluster, the interior points still have 12 nearest neighbours, but the exterior points have fewer. By correcting for the bonds lost from faces, edges and vertices, we obtain (\ref{shortbonds}). Similarly, in the infinite FCC lattice, each point has 6 next-to-nearest neighbours, so there are 3 long bonds per baryon. In a finite cluster, we find the number of long bonds is
\bea
N_{\rm long}(P,Q,R) &=& 2Q^3 + Q + \left(6Q^2 - \frac{1}{2}\right)(P+R) 
+ 3Q(P+R)^2 \nonumber \\
&& \ + (12Q + 3)PR + \frac{1}{2}(P+R)^3 + 3(P+R)PR \,.
\label{longbonds}
\eea

\section{Weight Clusters as Skyrmions}

To convert a weight cluster into a stable Skyrmion, nearest neighbour $B=1$ Skyrmions should have a relative orientation that minimises the energy of the pair, so there is maximal attraction. It is known how to assign orientations in an infinite FCC lattice, so that there is maximal attraction between each Skyrmion and all its 12 nearest neighbours. Four orientations are needed, arranged periodically. These are specified by four SO(3) matrices, but it is convenient to replace them by the four quaternions $1,i,j,k$. (This involves a sign choice that is unimportant here, but becomes more significant when we consider quantization.) Skyrmions which lie at points of the form $(0,0,0)$ mod 4 have orientation $1$, and points $(0,2,2)$, $(2,0,2)$ and $(2,2,0)$ mod 4 have, respectively, orientations $i$, $j$ and $k$.

These orientations are copied on to any weight cluster of quadrality 0, which is a subcluster of the FCC lattice. For the other quadralities, the cluster has to be translated, so that the weights lie on the same FCC lattice. Then the orientations can be copied directly. (We do not specify the translation precisely here; different choices correspond to different overall orientations, but in the quantization we have to consider all of these anyway.)

For example, the cluster with $c = 13$ and quadrality 0 becomes a $B=13$ Skyrmion consisting of a $B=1$ Skyrmion at the centre with orientation 1, surrounded by four $B=1$ Skyrmions with orientation $i$, four with orientation $j$ and four with orientation $k$. The $c = 4$ cluster has one $B=1$ Skyrmion of each orientation. In \cite{GHKMS}, the most strongly bound clusters up to baryon number 23 were listed (only a few of these have T or O symmetry), and in each case the distribution of orientations was given. It appears that the energy is usually lowest for clusters where the orientations are balanced (occur in equal or near-to-equal numbers). For quadrality 1 clusters with T symmetry, there is always exact balance, because each orbit of the group T with 24, 12 or 4 points is balanced in itself.

\begin{figure}
	\centering
	\begin{subfigure}{0.22\textwidth}
		\includegraphics[width=\textwidth]{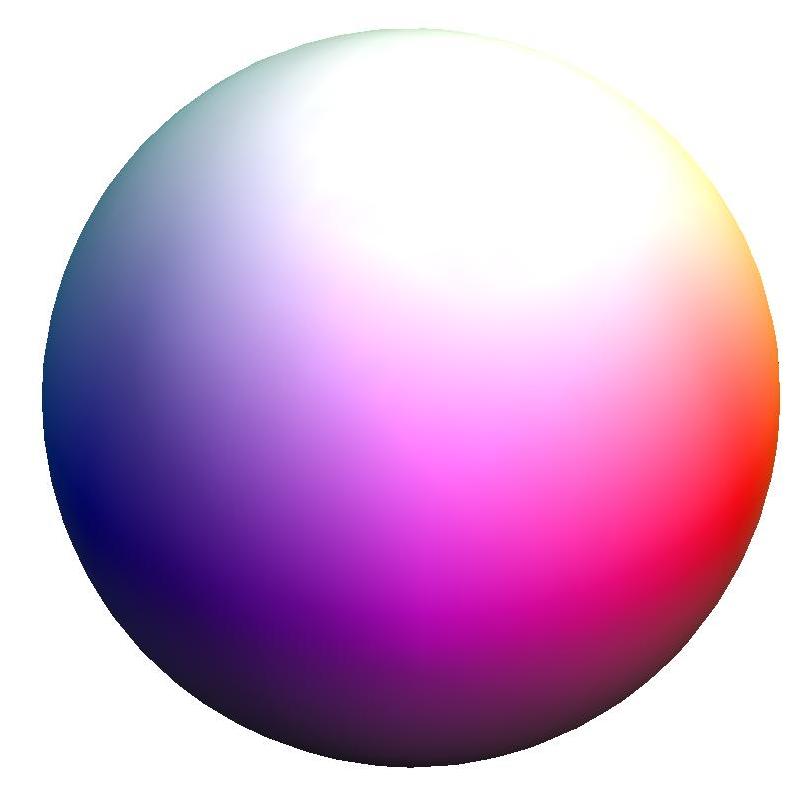} 
		\caption*{$B=1$}
	\end{subfigure}
	\begin{subfigure}{0.22\textwidth}
		\includegraphics[width=\textwidth]{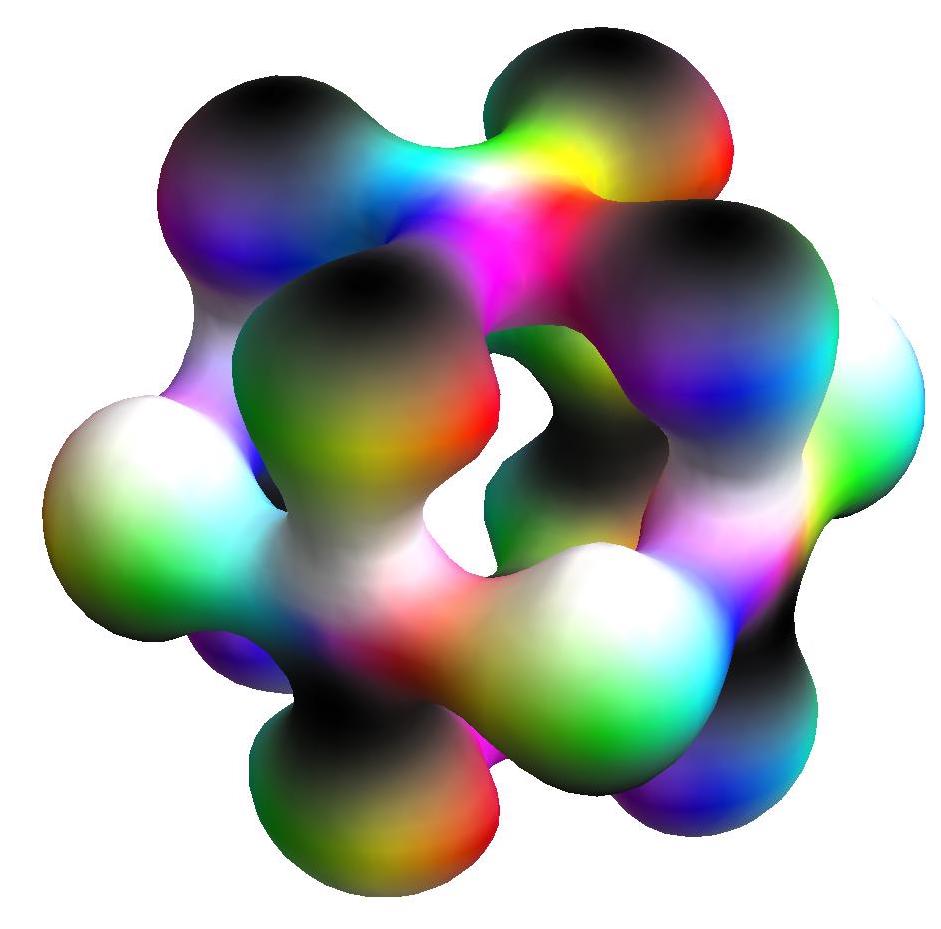} 
		\caption*{$B=13$}
	\end{subfigure}
	\begin{subfigure}{0.22\textwidth}
		\includegraphics[width=\textwidth]{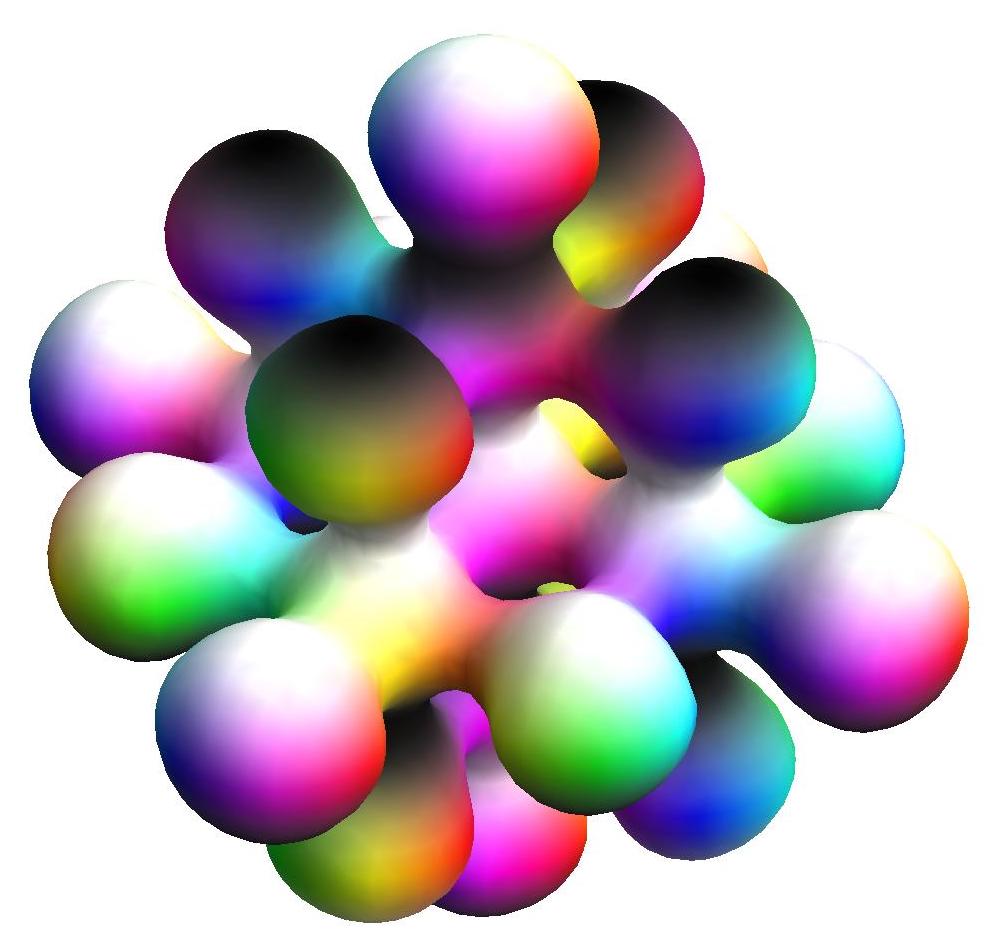} 
		\caption*{$B=19$}
	\end{subfigure}
	\begin{subfigure}{0.22\textwidth}
		\includegraphics[width=\textwidth]{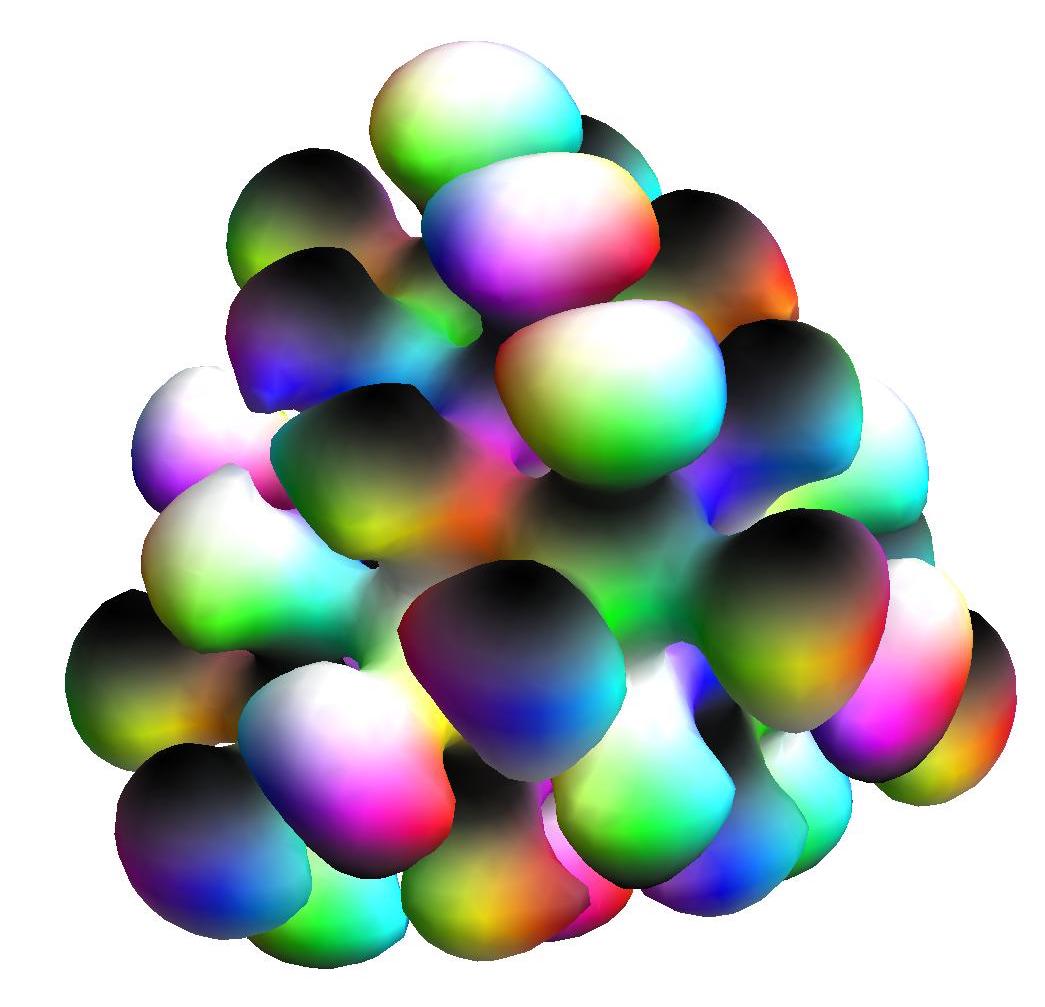} 
		\caption*{$B=31$}
	\end{subfigure}
	
	\begin{subfigure}{0.22\textwidth}
	\includegraphics[width=\textwidth]{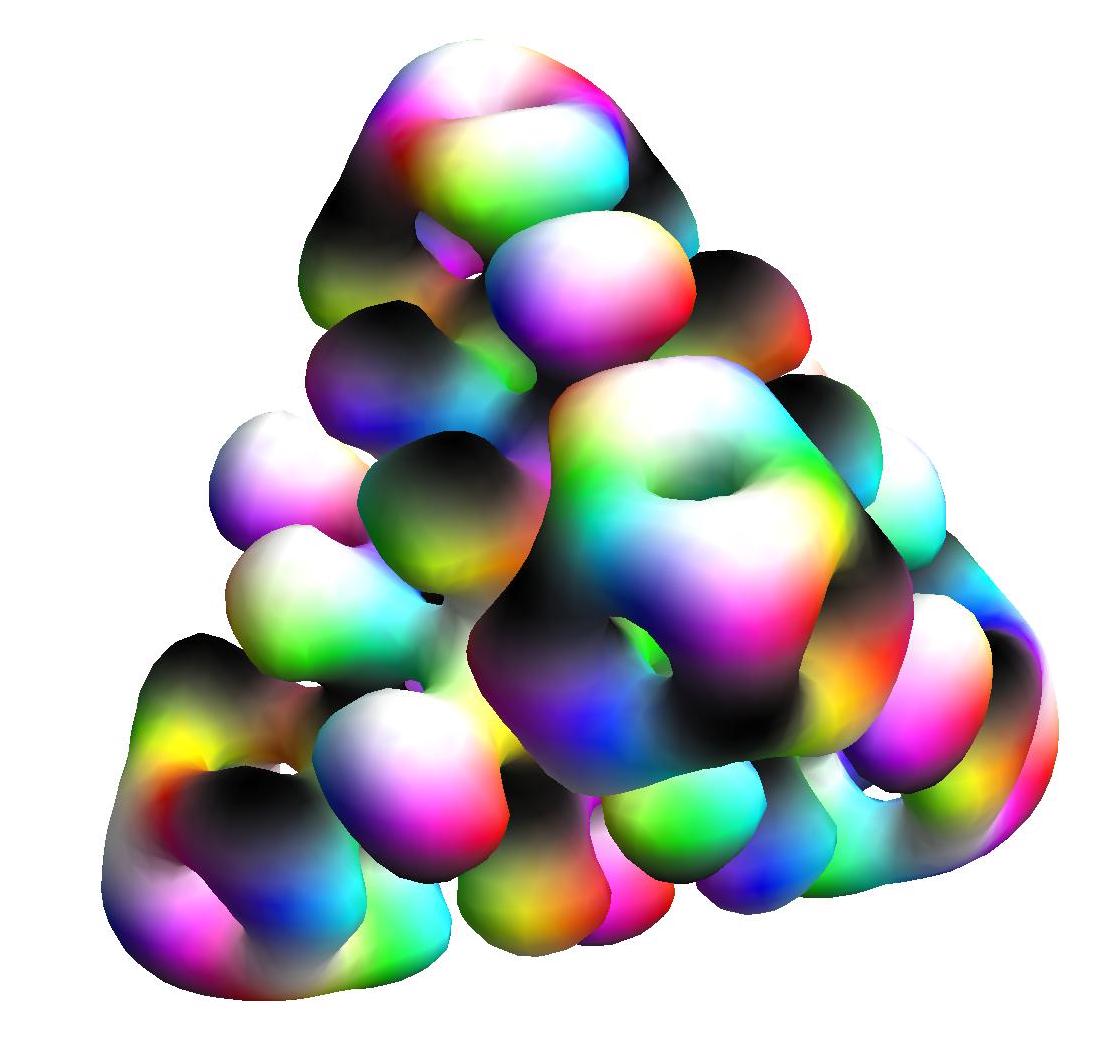} 
	\caption*{$B=35$}
\end{subfigure}
\begin{subfigure}{0.22\textwidth}
	\includegraphics[width=\textwidth]{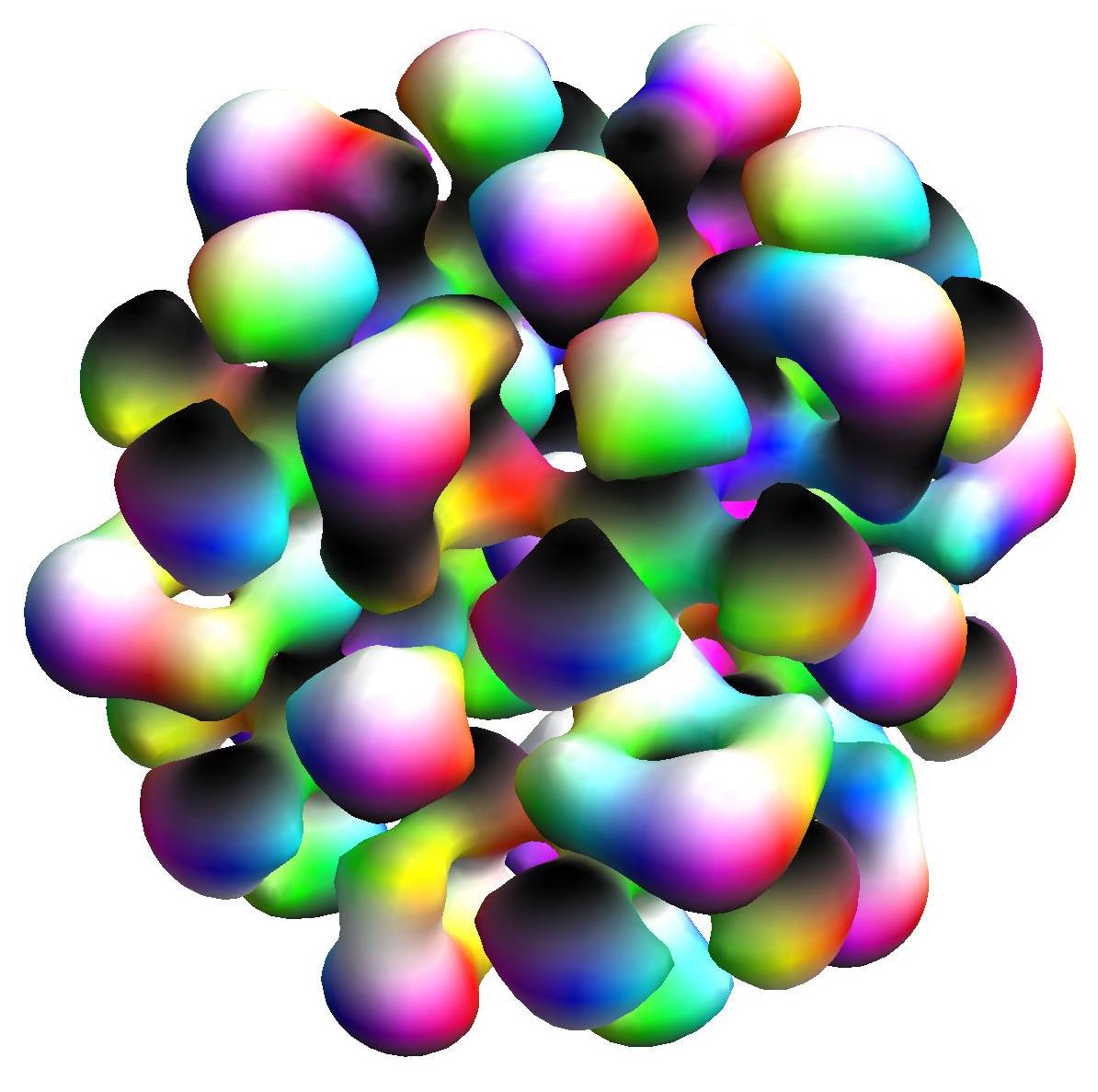} 
	\caption*{$B=55$}
\end{subfigure}
\begin{subfigure}{0.22\textwidth}
	\includegraphics[width=\textwidth]{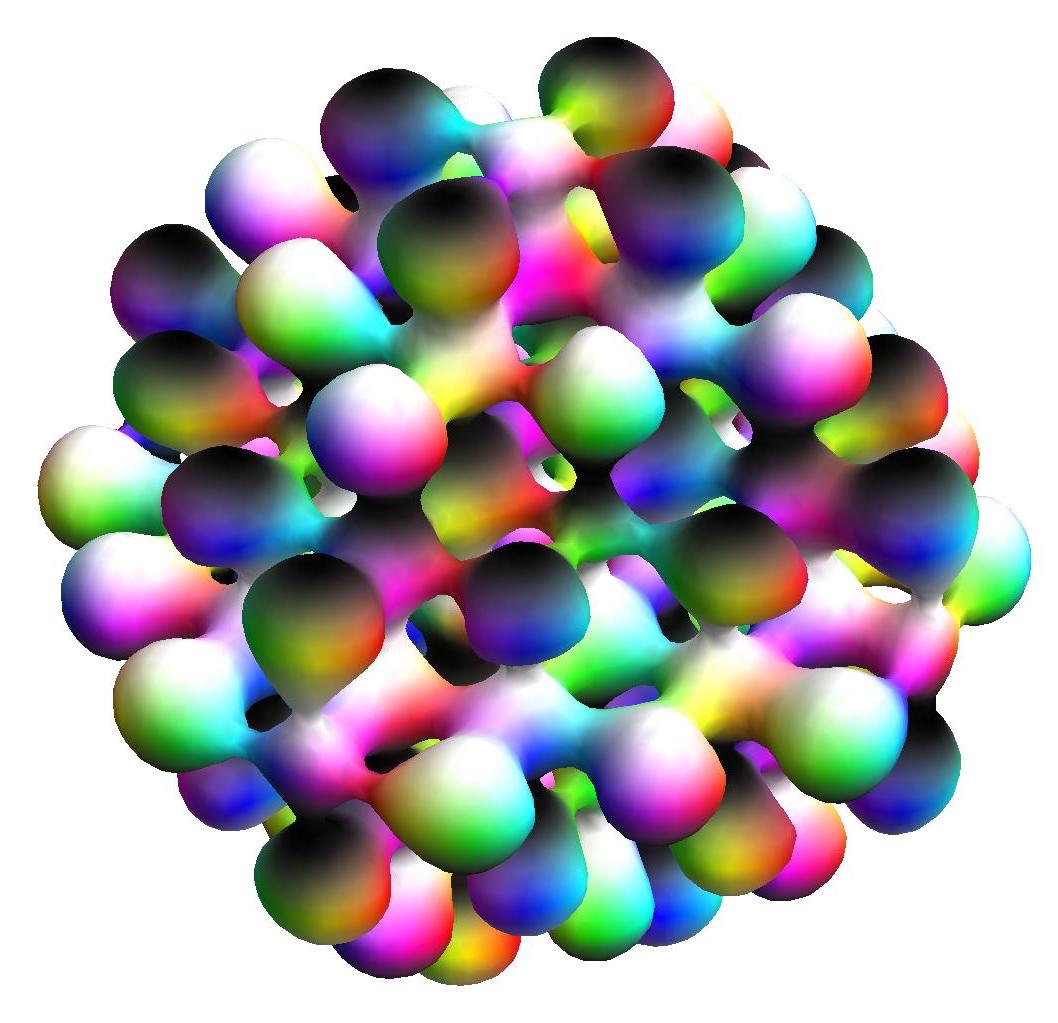} 
	\caption*{$B=79$}
\end{subfigure}
\begin{subfigure}{0.22\textwidth}
	\includegraphics[width=\textwidth]{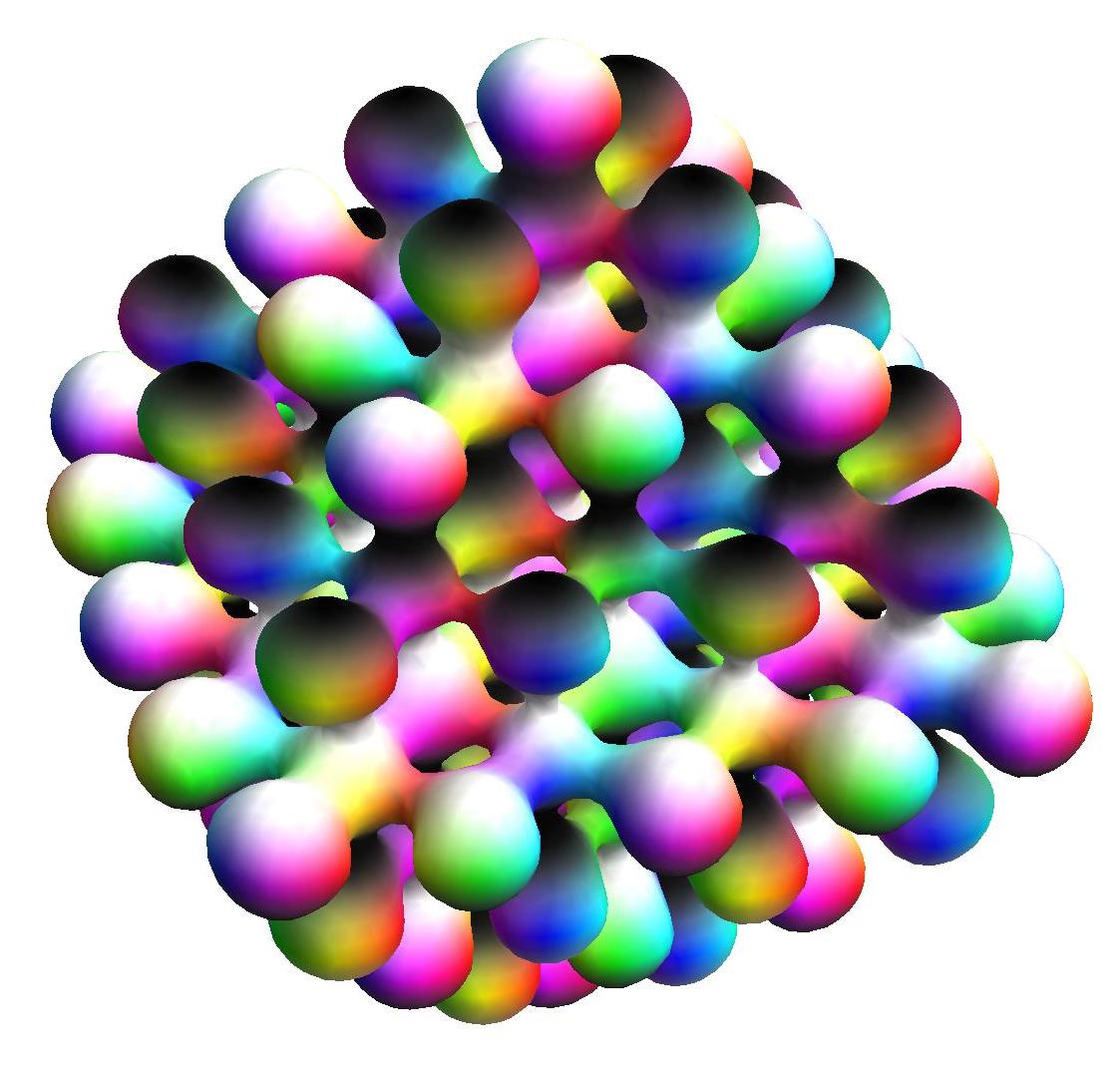} 
	\caption*{$B=85$}
\end{subfigure}
	
	\caption{The relaxed quadrality 0 Skyrmions.} \label{Q0sk}
	
\end{figure}

\begin{figure}
	\centering
	\begin{subfigure}{0.22\textwidth}
		\includegraphics[width=\textwidth]{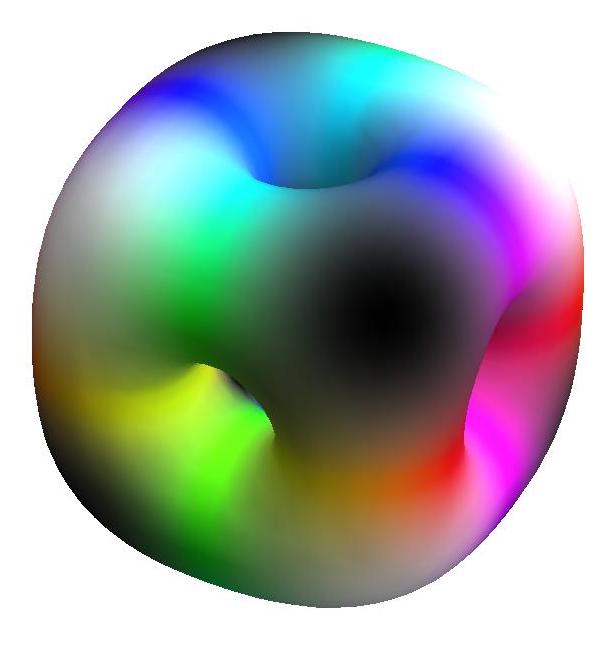} 
		\caption*{$B=4$}
	\end{subfigure}
	\begin{subfigure}{0.22\textwidth}
		\includegraphics[width=\textwidth]{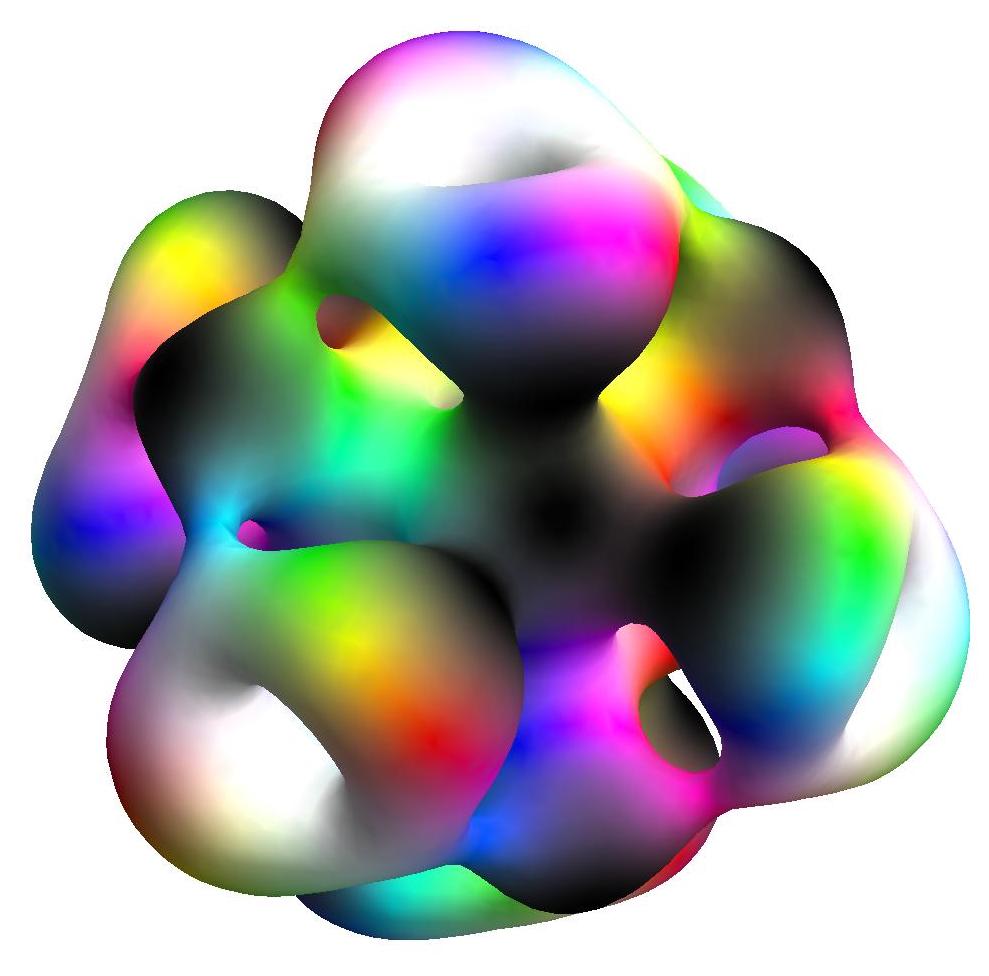} 
		\caption*{$B=16$}
	\end{subfigure}
	\begin{subfigure}{0.22\textwidth}
		\includegraphics[width=\textwidth]{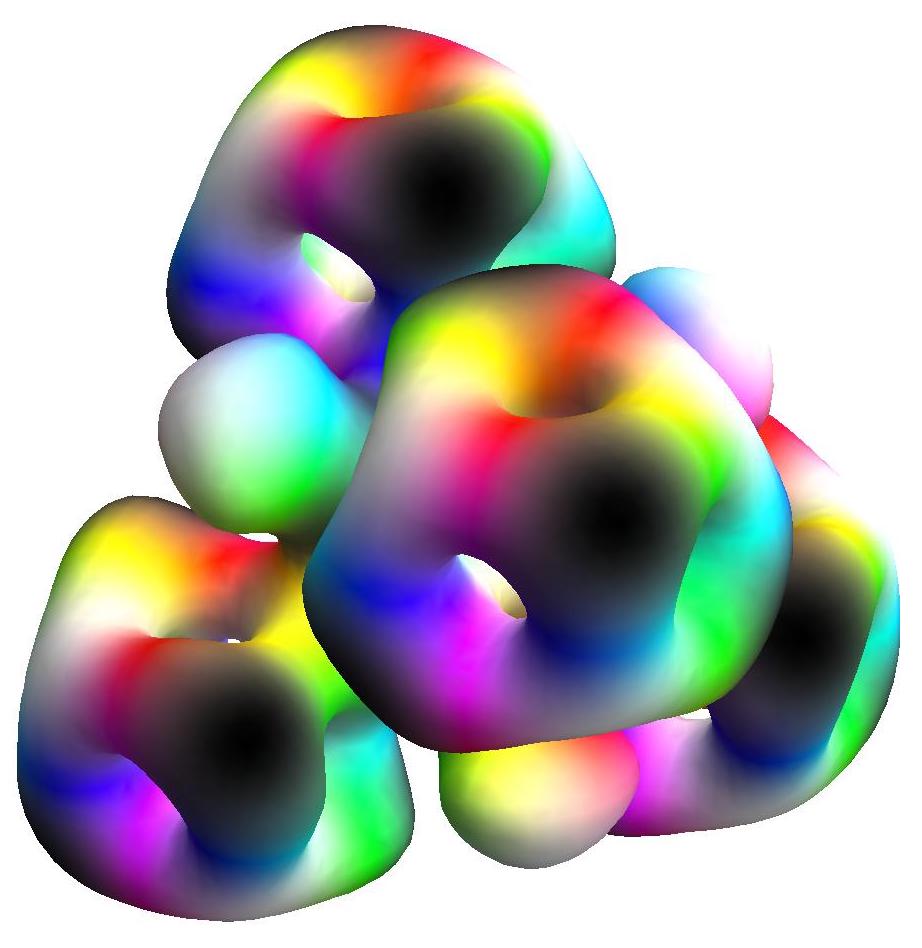} 
		\caption*{$B=20$}
	\end{subfigure}
	\begin{subfigure}{0.22\textwidth}
		\includegraphics[width=\textwidth]{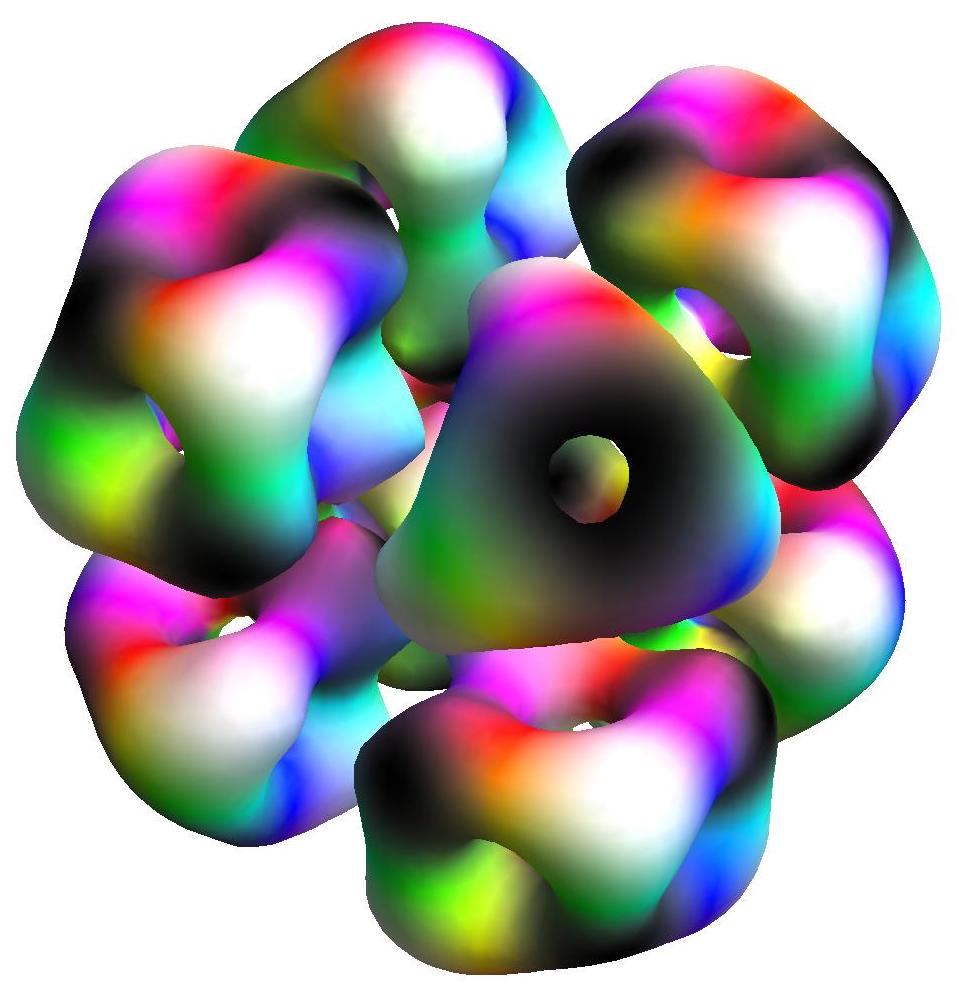} 
		\caption*{$B=28$}
	\end{subfigure}
	
	\begin{subfigure}{0.22\textwidth}
		\includegraphics[width=\textwidth]{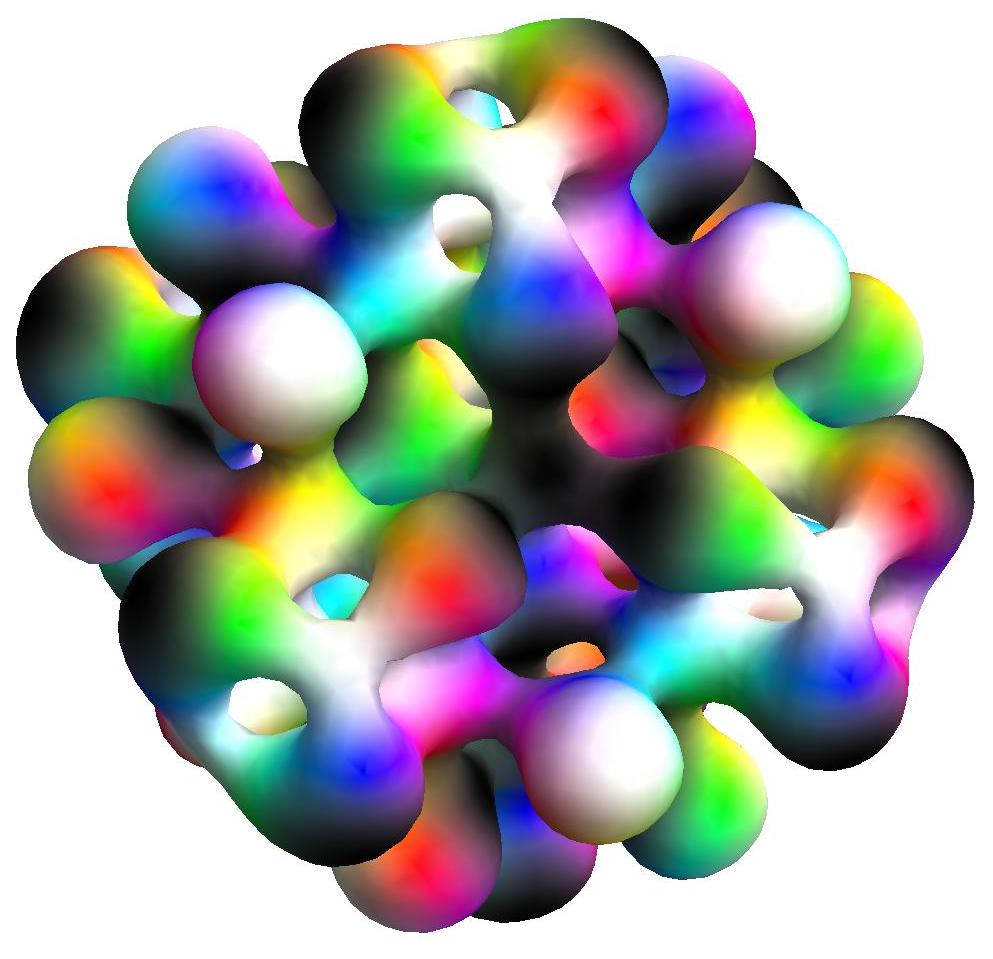} 
		\caption*{$B=40$}
	\end{subfigure}
	\begin{subfigure}{0.22\textwidth}
		\includegraphics[width=\textwidth]{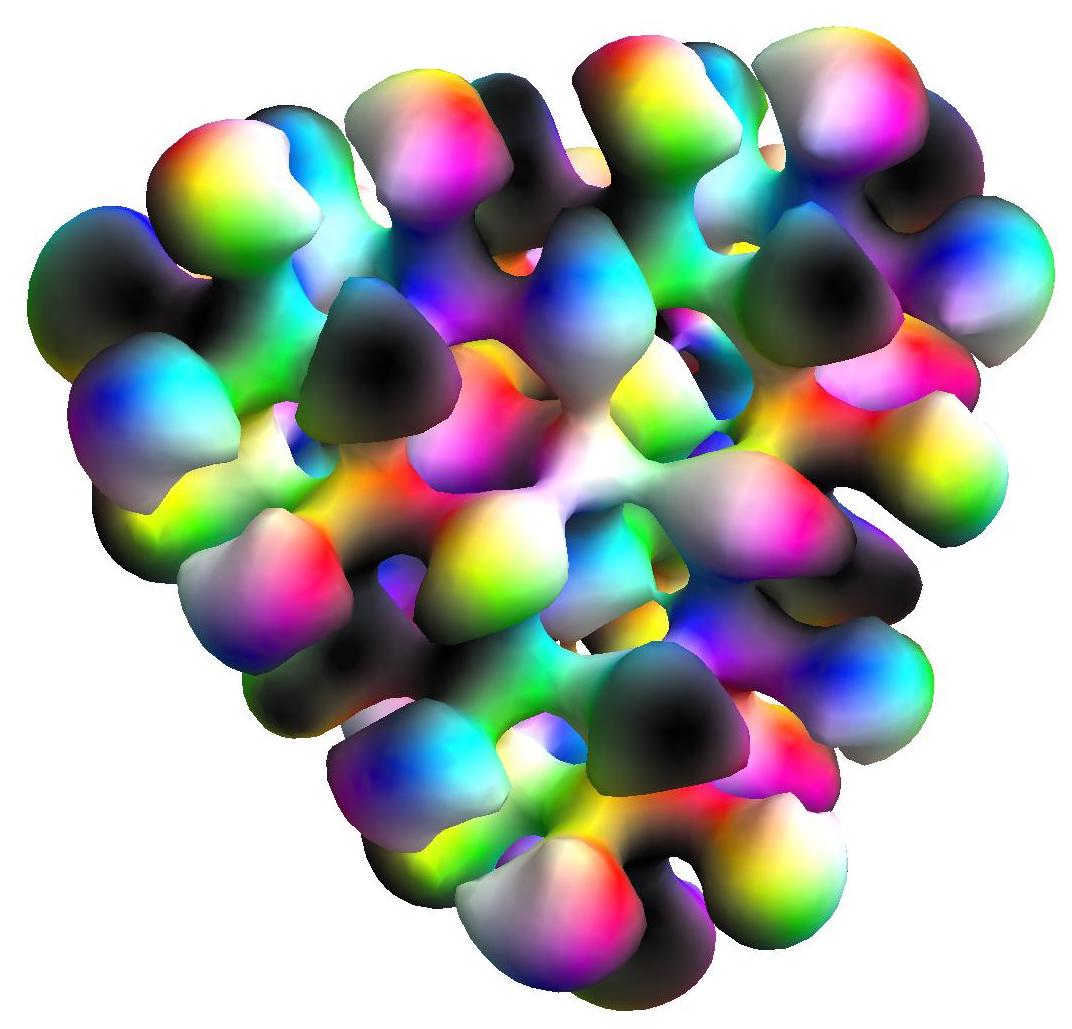} 
		\caption*{$B=52$}
	\end{subfigure}
	\begin{subfigure}{0.22\textwidth}
		\includegraphics[width=\textwidth]{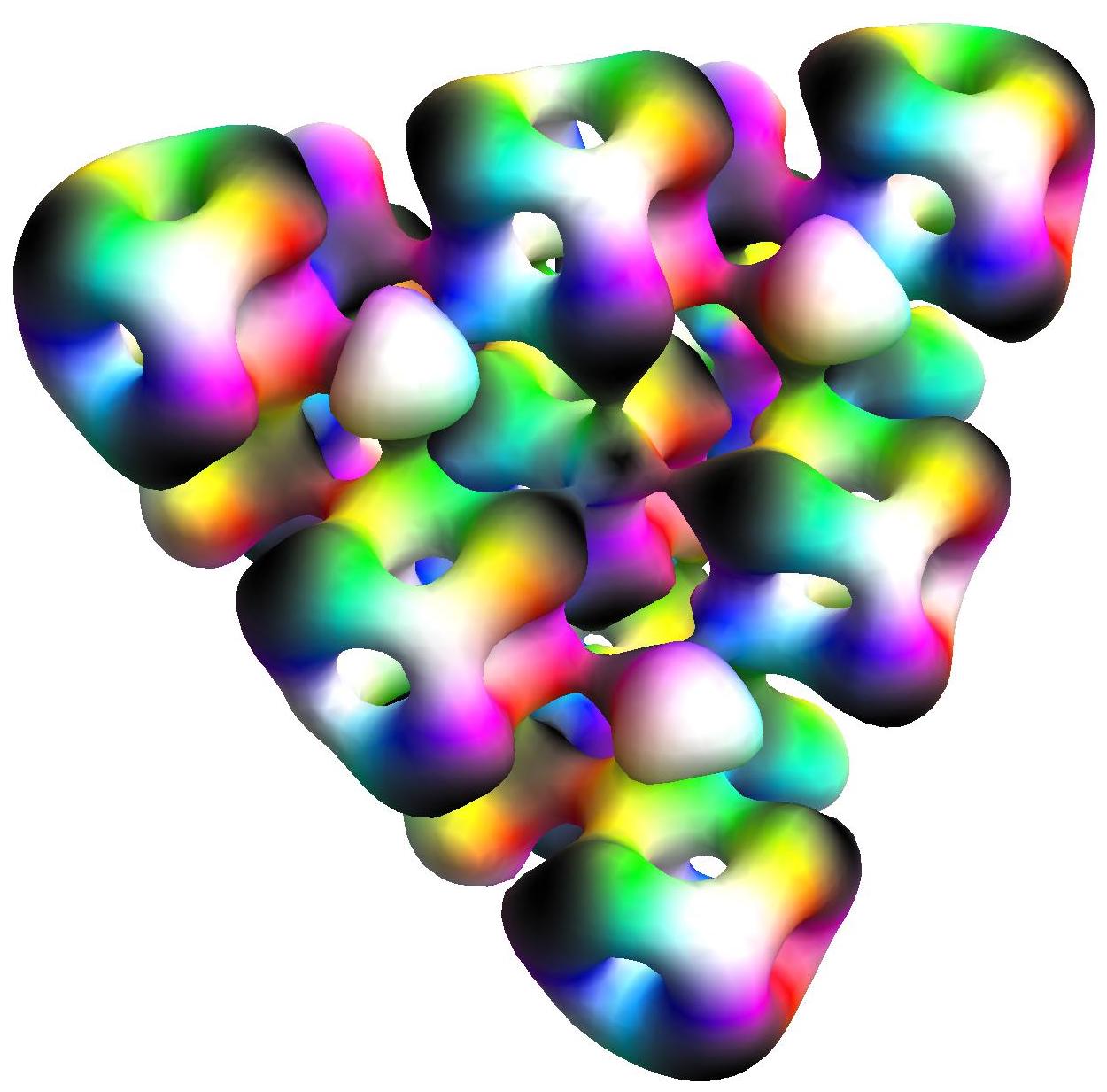} 
		\caption*{$B=56$}
	\end{subfigure}
	\begin{subfigure}{0.22\textwidth}
		\includegraphics[width=\textwidth]{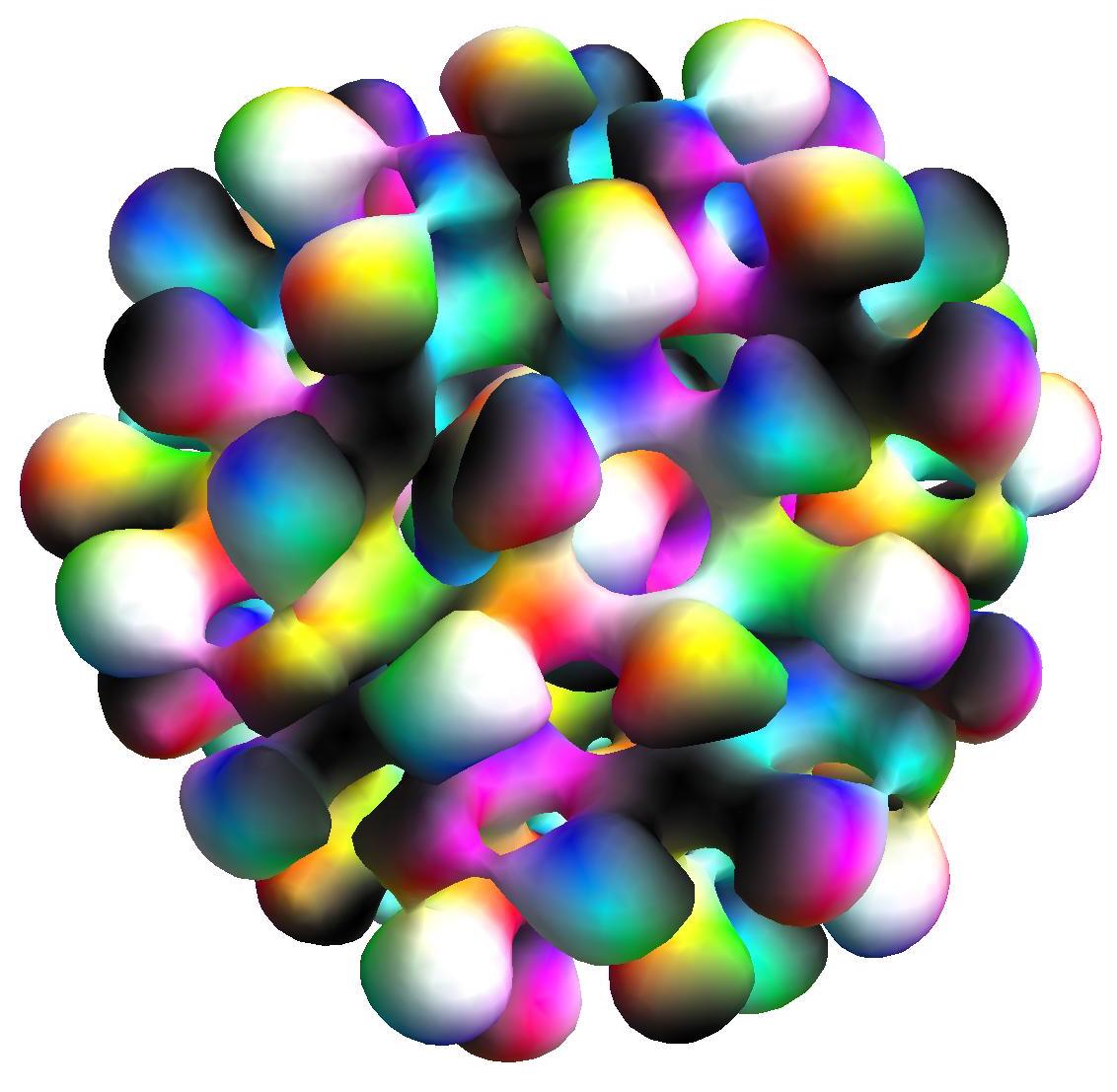} 
		\caption*{$B=68_1$}
	\end{subfigure}

	\begin{subfigure}{0.22\textwidth}
	\includegraphics[width=\textwidth]{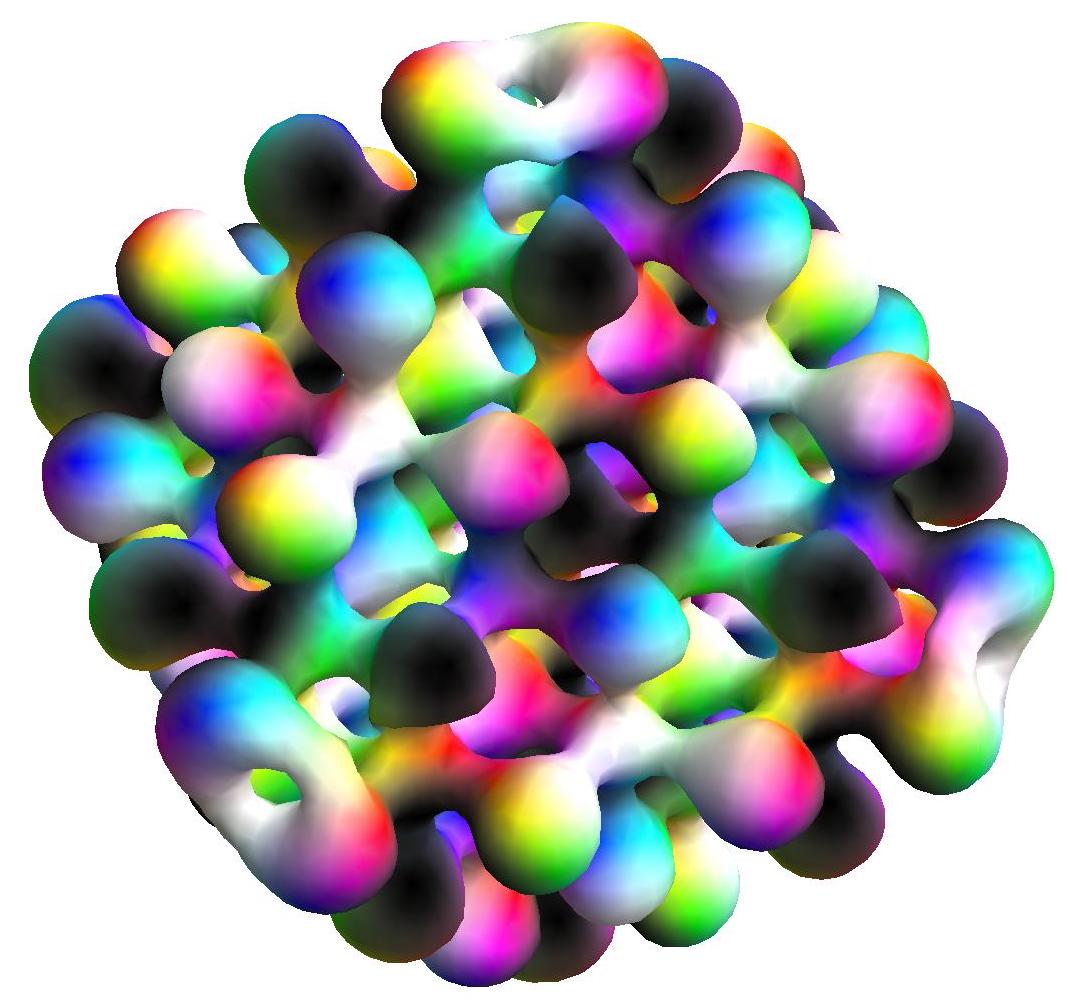} 
	\caption*{$B=80_{1a}$}
\end{subfigure}
\begin{subfigure}{0.22\textwidth}
	\includegraphics[width=\textwidth]{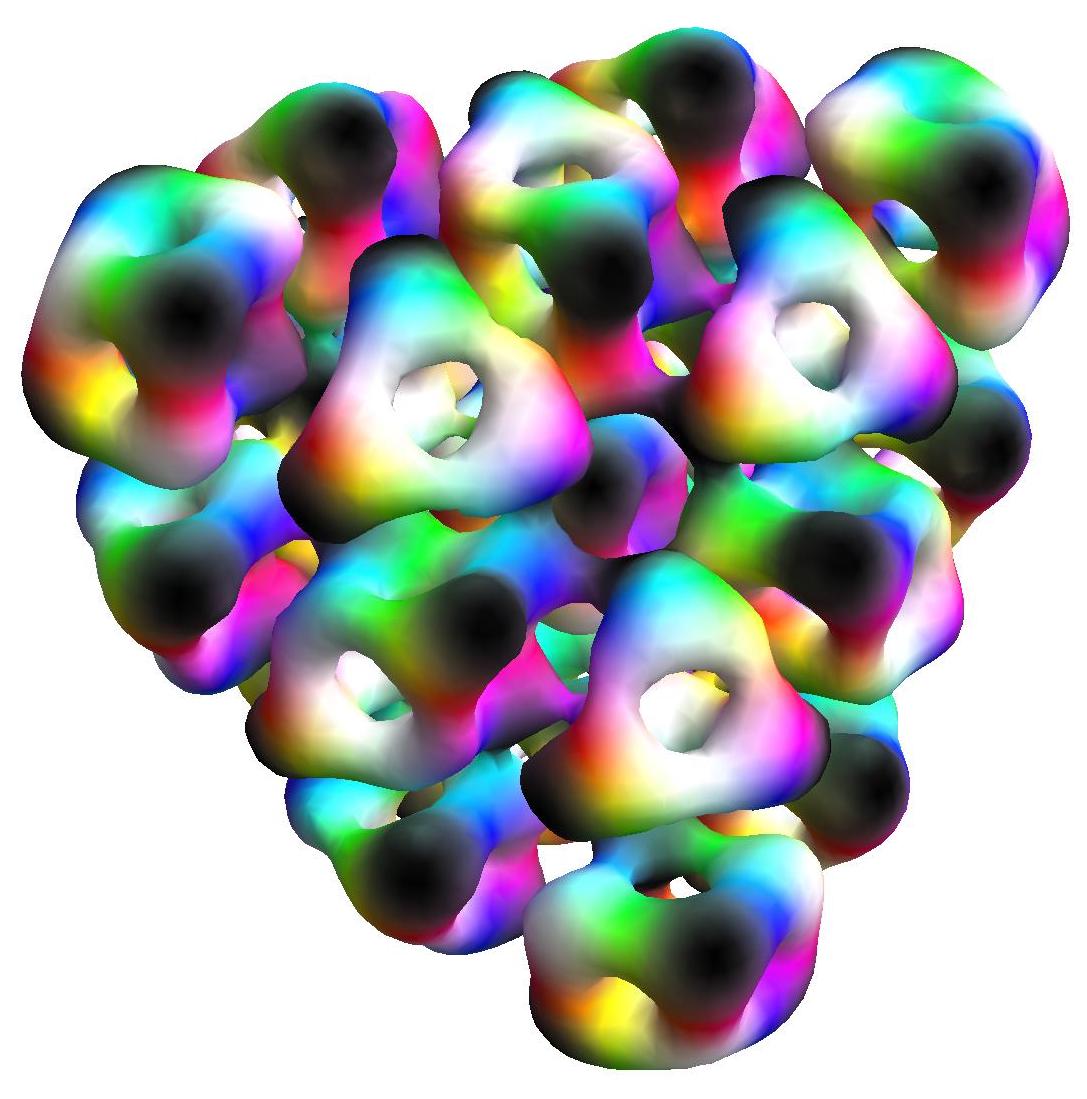} 
	\caption*{$B=80_{1b}$}
\end{subfigure}

	\caption{The relaxed quadrality 1 Skyrmions.} \label{Q1sk}
	
\end{figure}

\begin{figure}
	\centering
	\begin{subfigure}{0.22\textwidth}
		\includegraphics[width=\textwidth]{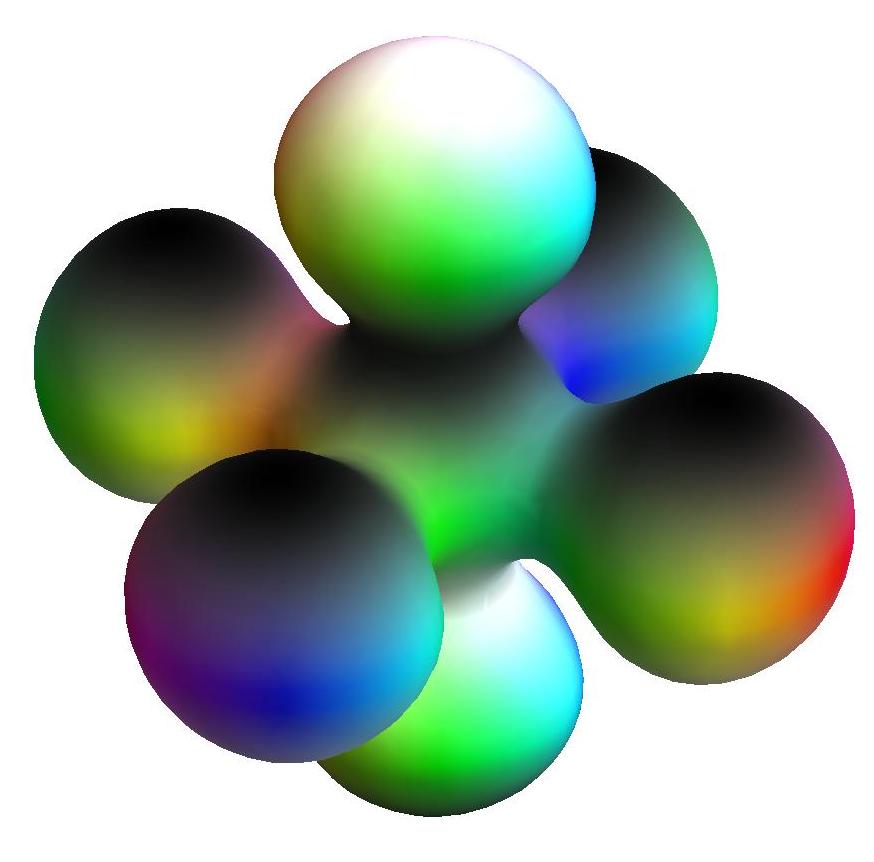} 
		\caption*{$B=6$}
	\end{subfigure}
	\begin{subfigure}{0.22\textwidth}
		\includegraphics[width=\textwidth]{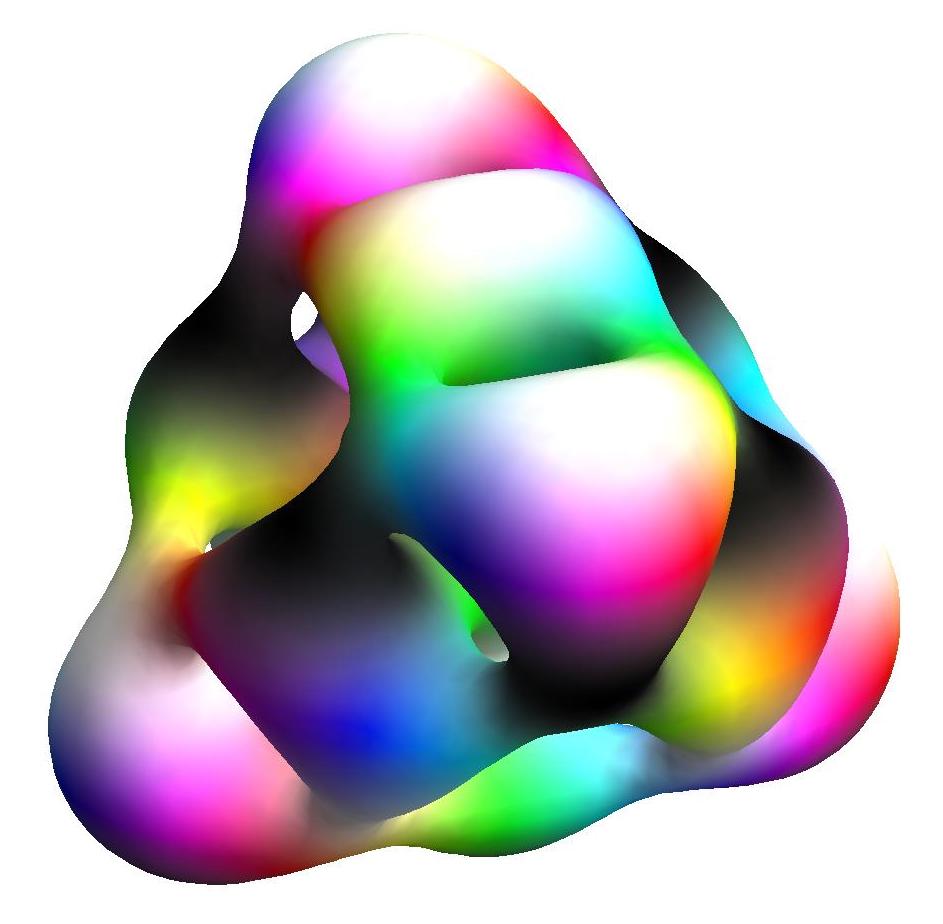} 
		\caption*{$B=10$}
	\end{subfigure}
	\begin{subfigure}{0.22\textwidth}
		\includegraphics[width=\textwidth]{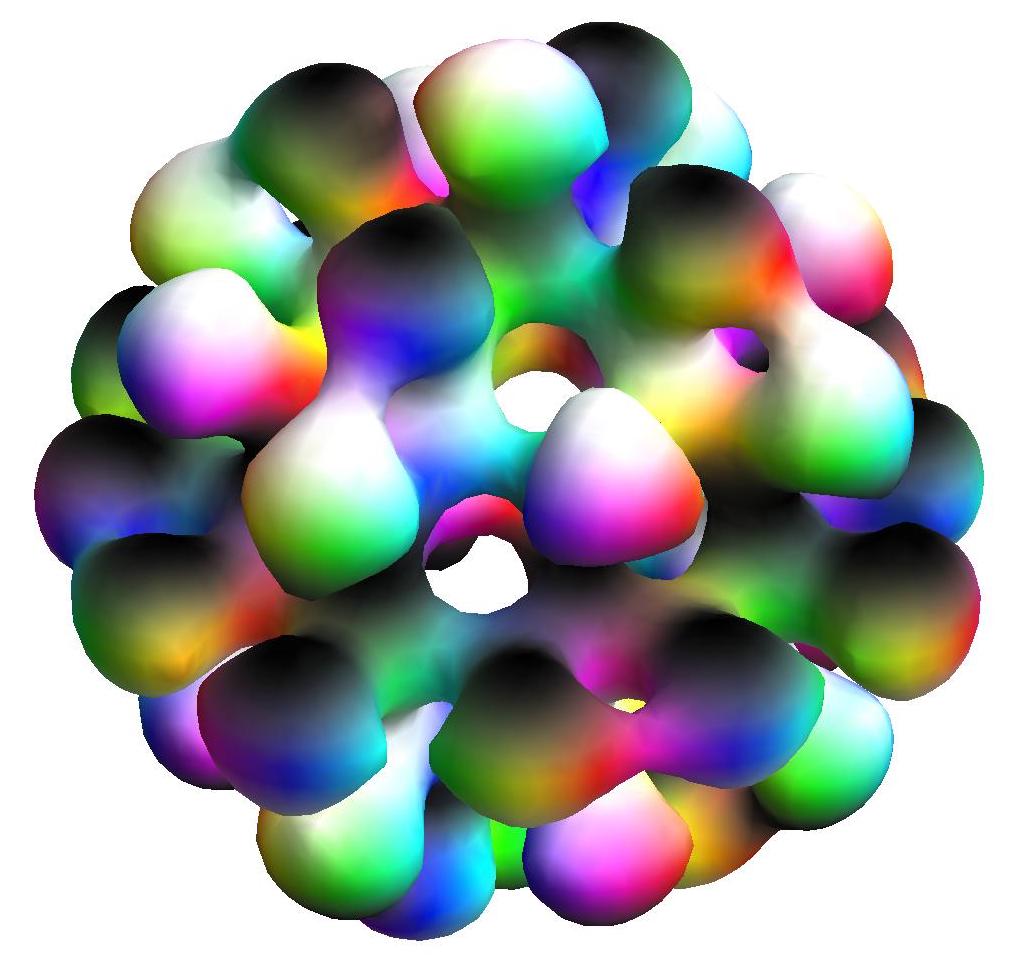} 
		\caption*{$B=38$}
	\end{subfigure}
	\begin{subfigure}{0.22\textwidth}
		\includegraphics[width=\textwidth]{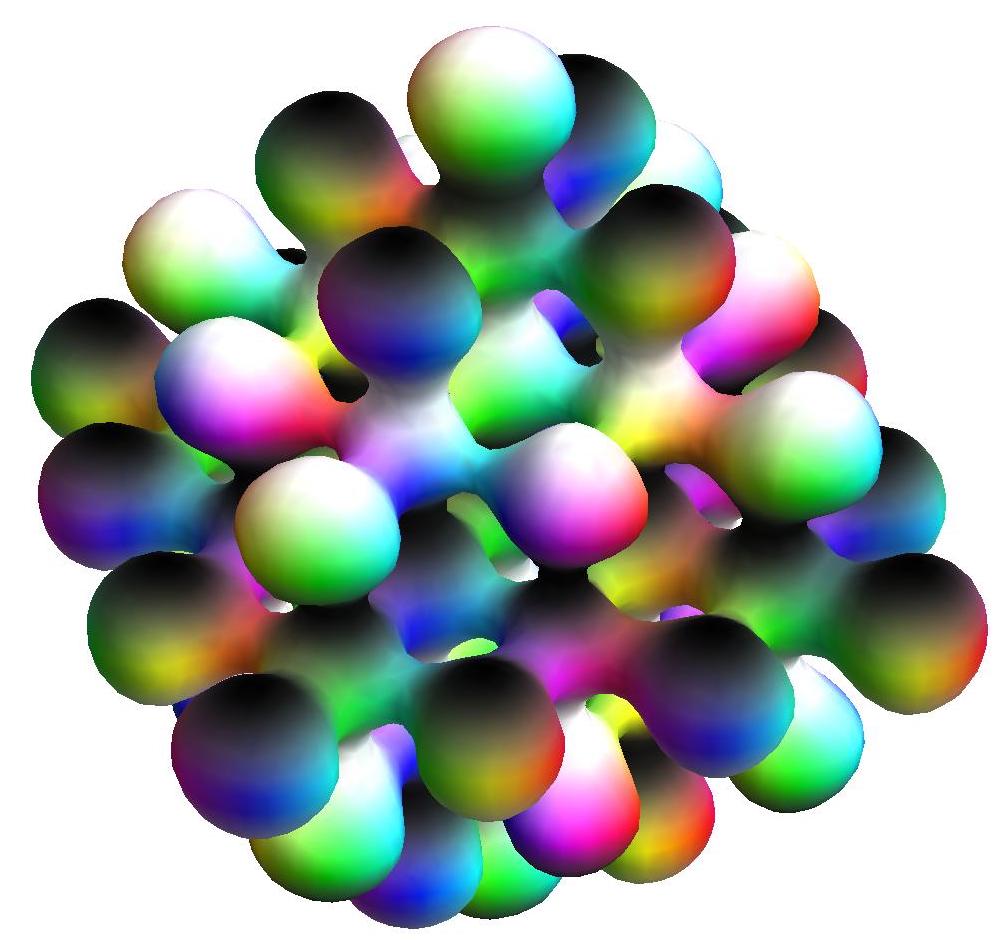} 
		\caption*{$B=44$}
	\end{subfigure}
	
	\begin{subfigure}{0.22\textwidth}
		\includegraphics[width=\textwidth]{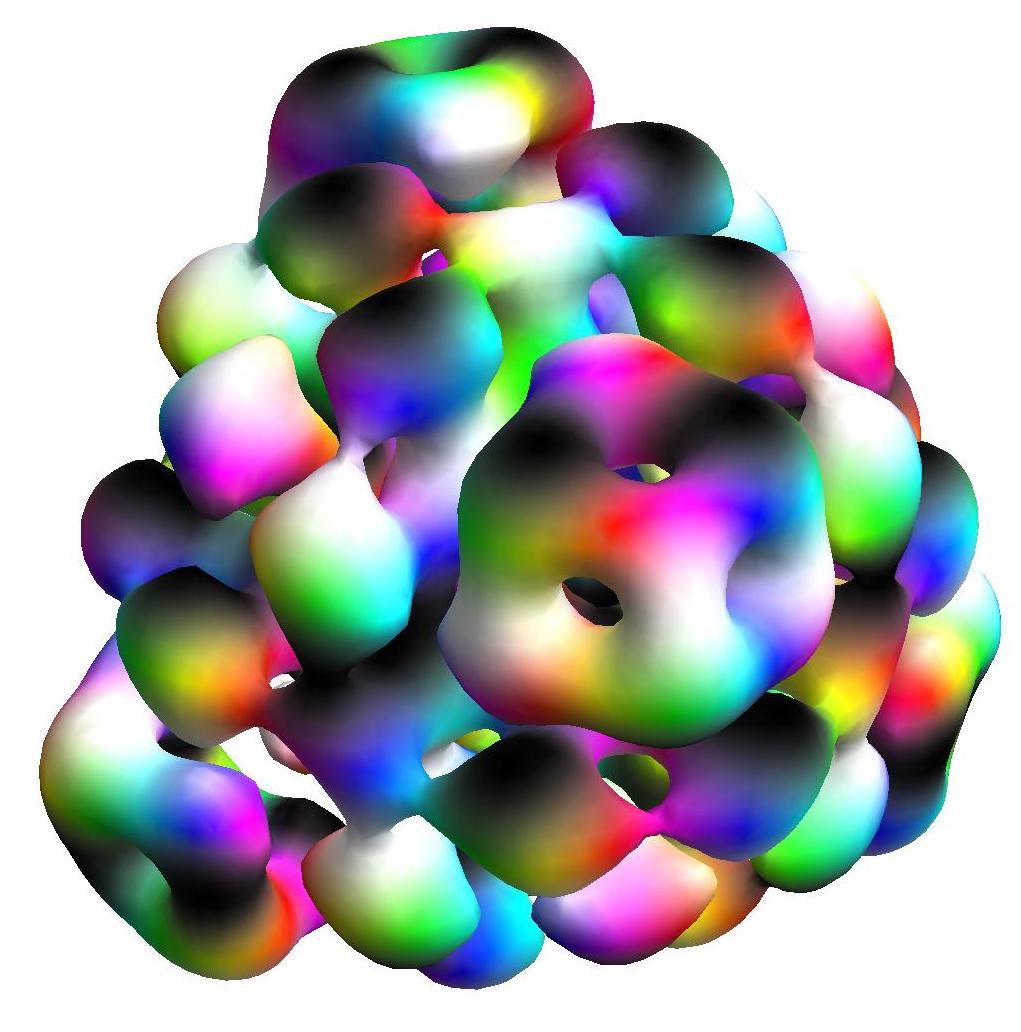} 
		\caption*{$B=50$}
	\end{subfigure}
	\begin{subfigure}{0.22\textwidth}
		\includegraphics[width=\textwidth]{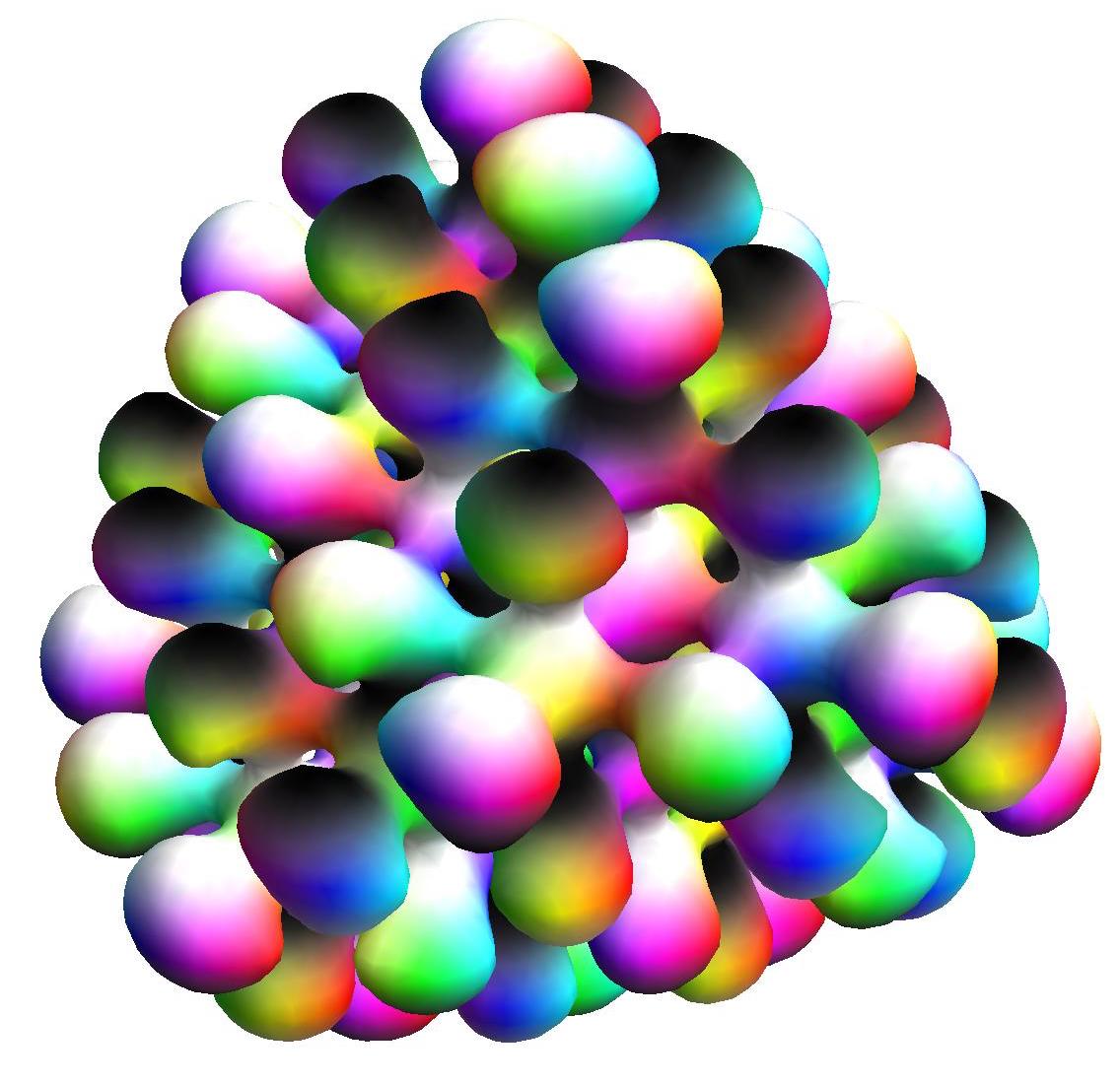} 
		\caption*{$B=68_2$}
	\end{subfigure}
	\begin{subfigure}{0.22\textwidth}
		\includegraphics[width=\textwidth]{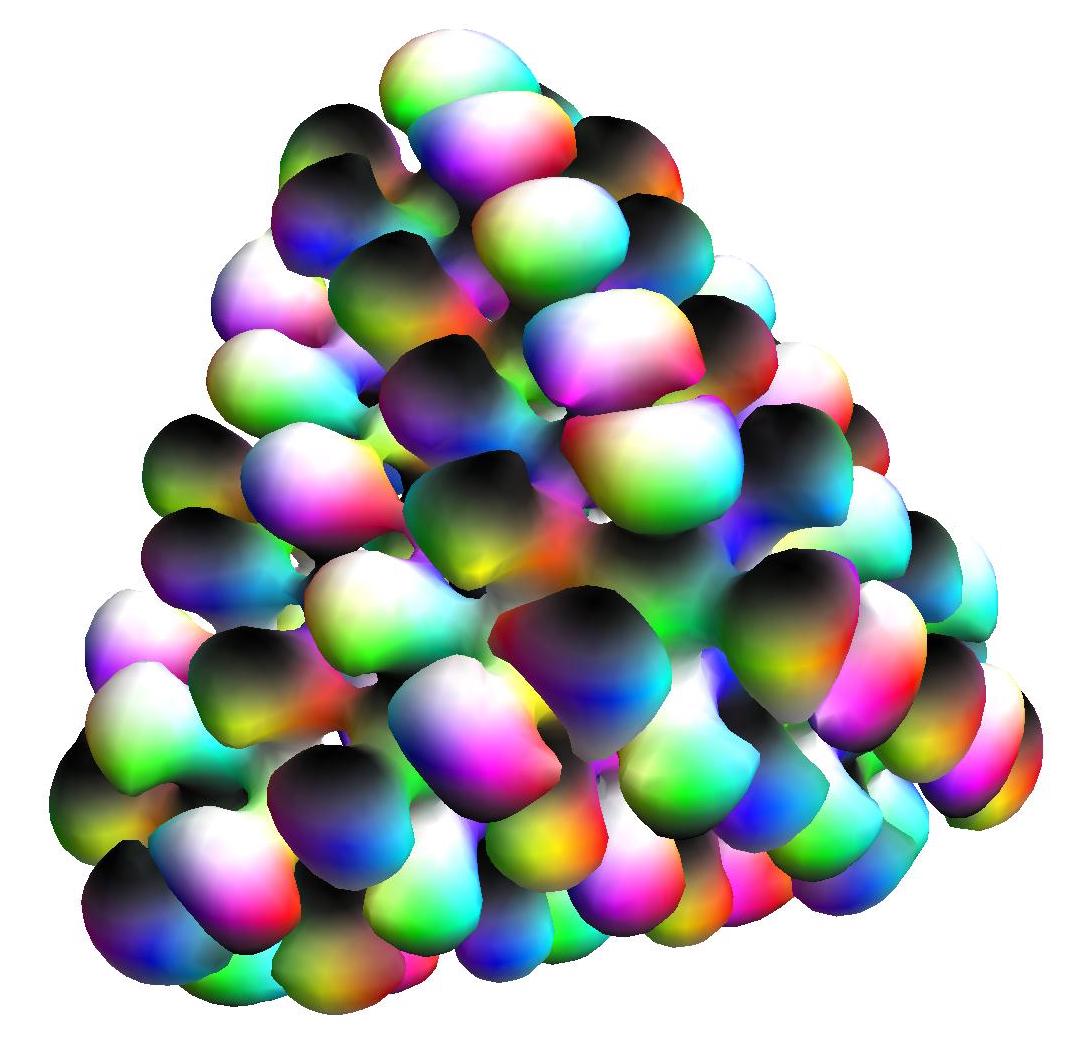} 
		\caption*{$B=80_{2}$}
	\end{subfigure}
	\begin{subfigure}{0.22\textwidth}
		\includegraphics[width=\textwidth]{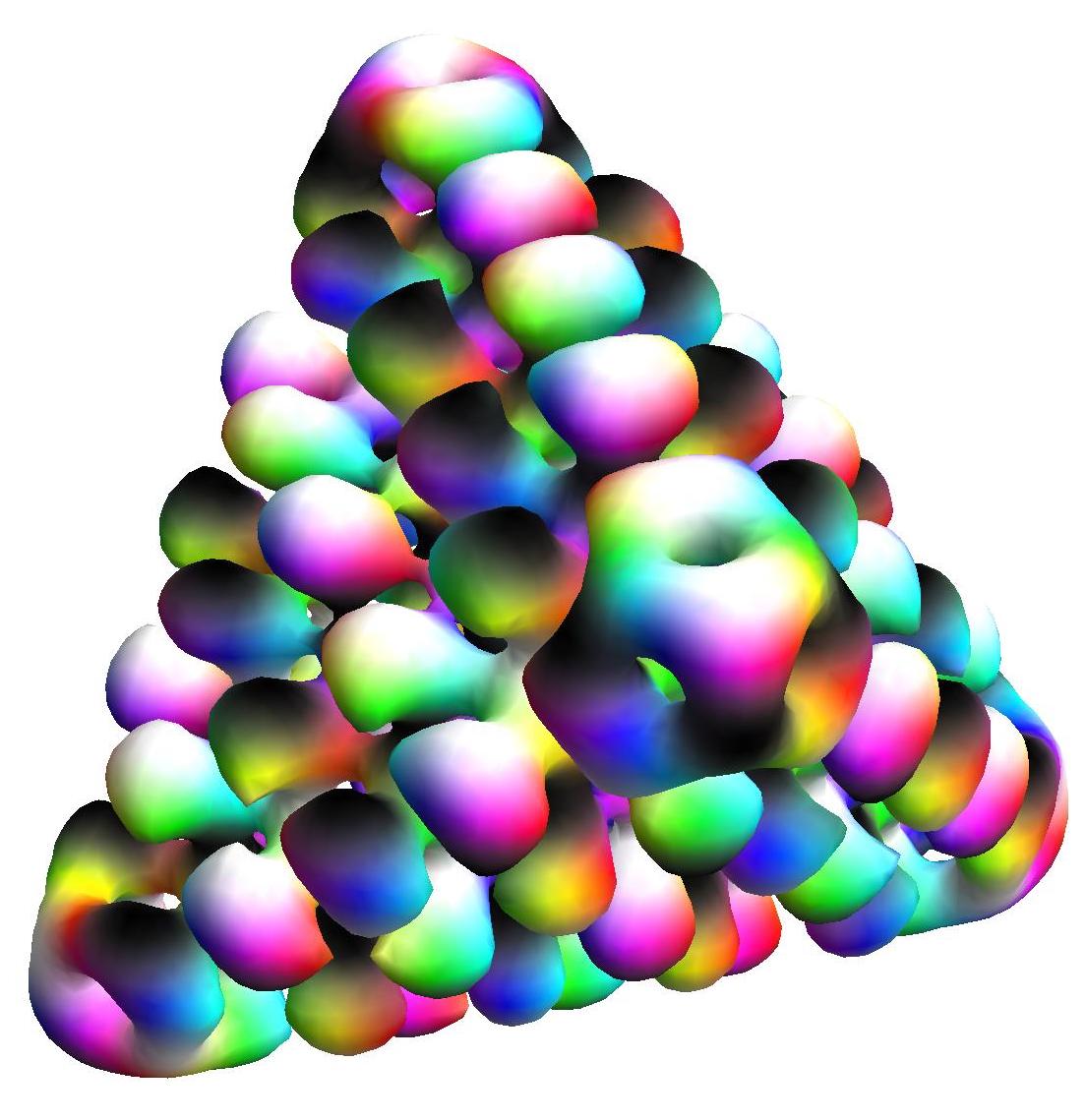} 
		\caption*{$B=84$}
	\end{subfigure}
	
	\caption{The relaxed quadrality 2 Skyrmions.}\label{Q2sk}
	
\end{figure}

The clusters of $B=1$ Skyrmions located at points of the FCC lattice are not true solutions of the standard Skyrme model, although they are close to being solutions of the lightly-bound model. We have found true solutions by relaxing the cluster configurations, using numerical gradient flow, and find that they are qualitatively similar to the clusters, but with some merging of the $B=1$ constituents. The numerically generated Skyrmions are presented in Figures \ref{Q0sk}, \ref{Q1sk} and \ref{Q2sk}. The symmetry of clusters is preserved by the relaxation and in one case is enhanced. The $B=4$ tetrahedral cluster relaxes to the well known, cubic $B=4$ Skyrmion with O symmetry, modelling the intrinsic structure of an alpha particle. Approximately cubic $B=4$ clusters appear in several larger $B$ clusters, e.g. in those with $B=20$ and $B=56$.

The properties of these numerically generated Skyrmions are presented in Table \ref{NumericalTable} alongside the estimate of the binding energy per baryon discussed in Section 2. We also list the moments of inertia of each Skyrmion. We follow \cite{MMW} by denoting the inertia tensors associated with angular motion, isoangular motion and their mixing as $V$, $U$ and $W$ respectively. The symmetry of each configuration restricts the form of the inertia tensors. For quadrality 0 and 2, each tensor is proportional to the unit matrix so that
\begin{equation}
V_{ij} = v \delta_{ij}, \, \, W_{ij} = w \delta_{ij}\,\text{ and }\, U_{ij} = u \delta_{ij}\, .
\end{equation}
Quadrality 1 Skyrmions have an additional independent component in the diagonalised $U$ matrix. We orient the Skyrmions so that $U_{11}=U_{22}$ and $U_{33}$ is independent. The difference between the two cases is due to the way the symmetry acts on the Skyrmion, which is discussed in Section 4.

\begin{figure}
	\centering
	\includegraphics[width=0.6\textwidth]{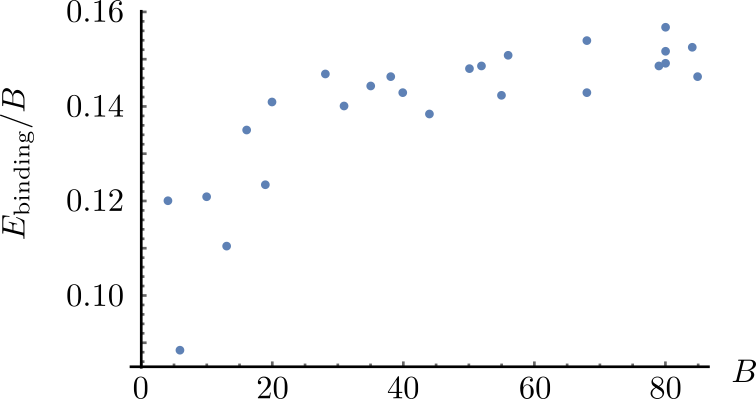} 
	\caption{The binding energy per baryon, in Skyrme units, for the numerically generated Skyrmions. We do not plot the result for the $B=1$ solution, which has zero classical binding energy.} \label{EnergyPerBaryon}
\end{figure}

The binding energy per baryon, $E_1 - E_B/B$, is plotted against baryon number in Figure \ref{EnergyPerBaryon}. Solutions with high binding energy are most likely to be relevant for nuclear physics. Generally, these have quadrality 1 and are balanced -- having equal numbers of the different $B=1$ orientations in the initial cluster. Skyrmions which are unbalanced, such as those with $B=55$ and $B=68_2$, have very low binding energy. The balanced configurations are more likely to show merging. For example, the tightly bound $B=56$ solution has the structure of ten $B=4$ Skyrmions, bound together by the remaining $B=1$ Skyrmions. The $B=28$ Skyrmion has a very large binding and exhibits the most obvious merging. It looks like four $B=4$ Skyrmions and four $B=3$ Skyrmions, with each set arranged tetrahedrally, and locked together. There are some exceptions to this rule. The $B=38$ and $B=52$ Skyrmions do not merge significantly but have high binding energy. The Skyrmions with small binding, for example, most of the quadrality $0$ clusters, usually do not merge significantly. For instance, one can clearly see the individual components of the $B=85$ solution. This has a binding energy of $0.146$ per baryon which is particularly low for its size, even less than the much smaller $B=28$ Skyrmion.

\begin{table}
	\begin{center}
		\begin{tabular}{| c | c | c | c | c | c | c | c | c | c | } \hline
			$B$ & Q & Sym & $E_B$ & $E_{\rm binding}/B$ &
            $E_{\rm bond}/B$ & $v$ & $w$ & $U_{11}$ & $U_{33}$  \\ \hline  
			1 & 0 & $O(3)$ &1.415 & 0 & 0 & 48 & 48 & 48 & 48 \\
			4 & 1 & O & 5.18 & 0.120 & 0.0780 & 661 & 0 & 147 & 176\\
			6 & 2 &  O & 7.96 & 0.0883 & 0.0832 & 1540 & -166 & 259 & 259  \\
			10 & 2 & T & 12.94 & 0.121 & 0.112 & 3750 & 145 & 398 & 398 \\
			13 & 0 & O & 16.96 & 0.110 & 0.106 & 6210 & -71 & 516&516  \\ 
			16 & 1 & T & 20.48 & 0.135 & 0.125 & 8230 & 0 & 672 & 656 \\
			19 & 0 & O & 24.53 & 0.123 & 0.125 & 11100 & 161 & 793 & 793 \\
			20 & 1 & T & 25.47 & 0.141 & 0.131 & 12800 & 0 & 756 & 819  \\
			28 & 1 & T & 35.51 & 0.147 & 0.125 & 23100 & 0 & 1024 & 1146  \\
			31 & 0 & T & 39.52 & 0.140 & 0.141 & 25000 & -113 & 1230 & 1230 \\
			35 & 0 & T & 44.47 & 0.144 & 0.143 & 33900 & 206 & 1380 & 1380 \\
			38 & 2 & O & 48.21 & 0.146 & 0.141 & 37500 & -115 & 1480 & 1480 \\
			40 & 1 & T & 50.88 & 0.143 & 0.145 & 38600 & 0 & 1600 & 1670 \\
			44 & 2 & O & 56.16 & 0.138 & 0.145 & 46900 & -268 & 1820 & 1820 \\
			50 & 2 & T & 63.50 & 0.148 & 0.137 & 64500 & 158 & 1890 & 1890 \\
			52 & 1 & T & 65.85 & 0.149 & 0.150 & 63400 & 0 & 2040 & 2070 \\
			55 & 0 & O & 70.01 & 0.142 & 0.136 & 74300 & 476 & 2160 & 2160 \\
			56 & 1 & T & 70.79 & 0.151 & 0.150 & 78200 & 0 & 2140 & 2270 \\
			$68_1$ & 1 & T & 85.70 & 0.154 & 0.150 & 99600 & 0 & 2660 & 2690 \\
			$68_2$ & 2 & T & 86.51& 0.143 & 0.154 & 99600 & 230 & 2730 & 2730 \\
			79 & 0 & O & 100.00& 0.149 & 0.155 & 125000 & 99 & 3190 & 3190  \\
			$80_{1a}$ & 1 & T & 101.27 & 0.149 & 0.146 & 129000 & 0 & 3280 & 3250 \\
			$80_{1b}$ & 1 & T & 100.66 & 0.157 & 0.156 & 139000 & 0 &  2900 & 3180 \\
			$80_{2}$ & 2 & T & 101.08 & 0.152 & 0.156 & 135000 & -52 & 3170 & 3170 \\
			84 & 2 & T & 106.00 & 0.153 & 0.156 & 156000 & 286 & 3320 & 3320 \\
			85 & 0 & O & 107.83 & 0.146 & 0.156 & 143000 & 252 & 3493 & 3493 \\
\hline
\end{tabular}
\vskip 7pt
\caption{The results of our numerical simulations. We tabulate the baryon number, quadrality, symmetry, energy, binding energy per baryon, estimated binding energy per baryon, and moments of inertia of each Skyrmion. The energies $E$ and $E_{\rm binding}/B$ are in Skyrme units, and the estimated binding energies per baryon $E_{\rm bond}/B$ are related to those in Tables \ref{Quad0table}, \ref{Quad1table} and \ref{Quad2table} by a conversion factor of $0.052$. For quadralities $0$ and $2$, $U_{11} = U_{33} = u$.}
\label{NumericalTable}
\end{center}
\end{table}

\section{Rigid Body Quantization}

The basic approach to quantization of a Skyrmion of any baryon number is the quantization of the overall orientational degrees of freedom in both space and isospace. This rigid body quantization is a generalisation of the treatment of the $B=1$ Skyrmion by Adkins, Nappi and Witten \cite{ANW}, and has been applied to many examples of Skyrmions, from $B=2$ upwards \cite{MMW, BC}. Shape-deforming degrees of freedom have also been considered more recently \cite{HKM, Hal}, but we will not consider them here. We will find the combinations of spin, isospin and parity that are allowed for the quantum states. A parity assignment emerges automatically whenever a Skyrmion has at least one reflection symmetry, and all our clusters have this.

Our clusters have either T or O symmetry, acting in the body-fixed frame of the Skyrmion. The even part of the symmetry group (the rotations in the tetrahedral or octahedral subgroups of SO(3)) places restrictions on the body-fixed spin and isospin quantum numbers, and the odd part controls the parity. If a given combination of total spin $J$ and total isospin $I$ is allowed, then their projections $J_3$ and $I_3$ on space-fixed axes are not constrained, and can take all their standard values ($2J+1$ values of $J_3$ and $2I+1$ values of $I_3$), so we suppress these labels when describing allowed states, and only keep the labels $J,L_3$ and $I,K_3$ where $L_3$ and $K_3$ are the projections of spin and isospin with respect to the third body-fixed axes. $K_3$ is a quantity peculiar to Skyrmions, where isospin arises from the quantization of the orientations in isospace. (Our notation is that usually used in the context of Skyrmion quantization, and also in particle physics. In nuclear physics, spin and isospin are usually denoted by $I$ and $T$.)
 
The symmetry groups T and O act on clusters by a combination of O(3) transformations (even or odd orthogonal transformations) in space and in isospace. For each spatial transformation $R$ there is a corresponding isospatial transformation $M(R)$, and the combined action leaves the classical Skyrmion invariant. The map $R \to M(R)$ is a representation of the symmetry group; this ensures the correct product structure for the combined actions. If $R$ is even/odd, then $M(R)$ is also even/odd. This is because a combination of even and odd O(3) transformations would reverse the sign of the baryon number $B$, and our clusters all have positive $B$.

The representation $R \to M(R)$ is the same for all the clusters with the same quadrality, and it can be determined by the symmetry group action close to the centre of a cluster. For quadrality 0 there is a single $B=1$ Skyrmion at the centre. The symmetry group, T or O, therefore acts in the same way as does the full symmetry group O(3) on a single $B=1$ Skyrmion, but restricted to the T or O subgroup. For the $B=1$ Skyrmion, the action of O(3) is with $M(R) = R$.
 
For quadrality 1 the symmetry is T and the central region contains a tetrahedral $B=4$ cluster. The way the tetrahedral group acts is well known from previous studies of the $B=16$ tetrahedrally symmetric Skyrmion. The mapping $R \to M(R)$ for the even part of T is one of the non-trivial 1-dimensional representations, i.e. the $\pi$ rotations are accompanied by trivial isorotations, and $\frac{2\pi}{3}$ rotations are accompanied by $\frac{2\pi}{3}$ isorotations in a particular plane in isospace. This is why there are two isorotational moments of inertia, $U_{11}$ and $U_{33}$. The $B=4$ cluster relaxes to a Skyrmion solution with enhanced octahedral symmetry, but this is not the case for larger tetrahedral clusters like $B=16$. So we shall ignore the enhanced symmetry, and the additional restrictions it places on allowed states.

For quadrality 2 the symmetry is either T or O, but there is always a $B=6$ octahedron at the centre. The fields produced by this central cluster have the same symmetry realisation as for quadrality 0, with $M(R) = R$. A way to understand this is to note that the FCC lattice of Skyrmions relaxes to a half-Skyrmion crystal with enhanced symmetry if the pion mass $m$ is zero. One may interpret the $B=6$ octahedron as approximately having a half-Skyrmion at its centre, with the same symmetry properties as the $B=1$ Skyrmion at the centre of a quadrality 0 cluster.

Knowing the way that the symmetry group T or O is realised for a cluster is not quite enough for writing down the rigid body quantization constraints. One needs to know in addition the Finkelstein--Rubinstein (FR) representation \cite{FR}. For each even symmetry operation there is an FR sign $\pm 1$. Collectively, these define a further representation of the even part of the symmetry group. The FR signs arise topologically, and ensure that individual $B=1$ Skyrmions are being quantized as fermions. Fundamentally, the FR sign is $+1$ or $-1$ depending on whether the loop generated by the symmetry operation is contractible or not in the configuration space of Skyrme fields. Here one extends a discrete symmetry to a closed loop by creating a path among rotations and isorotations connecting the initial cluster to the (identical) rotated cluster. However, determining directly the contractibility or non-contractibility of a loop is not trivial, and certain algorithms have been developed to get around this.

The even part of the tetrahedral group T has no non-trivial 1-dimensional representation involving just $+1$ and $-1$. The FR signs for the symmetry group T must therefore all be $+1$. For the even part of the octahedral group O, there is the trivial irrep $A_1$, where all signs are $+1$, but also the non-trivial 1-dimensional irrep $A_2$, where the rotations by $\frac{\pi}{2}$ around octahedral vertices and also rotations by $\pi$ around octahedral edge centres are represented by $-1$. There are therefore two possibilities for FR signs for the group O, which we denote by ${\rm O}+$ and ${\rm O}-$ respectively.

For octahedrally symmetric clusters of quadralities 0 and 2, both FR sign representations can occur, and we next use the algorithm presented in \cite{GHKMS} for finding which one does. The clusters are either pure octahedra or truncated octahedra, where all six vertices are cut off to leave square faces, and it is sufficient to calculate the FR sign associated to a $\frac{\pi}{2}$ rotation around one of these vertices. To do this, consider the cluster sliced  into planar layers orthogonal to the rotation axis. There is one central layer, and pairs of similar layers above and below this, with square symmetry. On each layer just two orientations of the $B=1$ Skyrmions occur, alternating in a chessboard pattern. The $\frac{\pi}{2}$ rotation has two types of orbit on cluster points -- either a single point, or a set of four points that are cyclically permuted.

Under the $\frac{\pi}{2}$ rotation permuting the points of the cluster, the changes in orientations that arise (assuming the $B=1$ Skyrmions just rotate around with the points) can be compensated by an overall isospin rotation. This is realised by conjugating all the orientation quaternions by a single (fixed) quaternion. This restores the SO(3) orientations, but each orientation quaternion may flip sign. The algorithm of \cite{GHKMS} is that the total FR sign is the product of the sign of the permutation and all the orientational sign flips.

We can arrange the cluster's initial orientation so that for the single-point orbits on the rotation axis, the rotation and the conjugation have a trivial effect. The orbits on the layers above and below the central layer have the same distribution of orientations, so their combined sign flip contribution is $+1$. For each four-point orbit in the central layer, the orientations occur in pairs, so the sign flips occur in pairs, again contributing $+1$. Only the contribution of the permutation of points remains. Now, a $\frac{\pi}{2}$ rotation is a cyclic permutation of a four-point orbit -- an odd permutation, contributing $-1$. Therefore, the FR sign is $(-1)^N$, where $N$ is the number of four-point orbits in the central layer. If this sign is positive, the FR representation of the symmetry group O is the trivial irrep $A_1$, and if it is negative the FR representation is the non-trivial irrep $A_2$. The number of four-point orbits is easily found, knowing the cluster shape. For the octahedral clusters with $B=1, 19, 44$ and $85$ it is even; for the octahedral cluster with $B=6$ and the truncated octahedra with $B=13, 38, 55$ and $79$ it is odd.

Some of these FR representations can be checked in other ways. For $B=1$, the FR representation is $A_1$ because the symmetry group is enhanced to a continuous group. For the $B=6$ cluster, the FR representation is $A_2$ because this cluster may be constructed approximately using a double rational map ansatz \cite{MP}, combining O-symmetric rational maps of degrees 1 and 5. Krusch's calculation \cite{Kru} shows that the FR sign associated with a $\frac{\pi}{2}$ rotation is $+1$ for the degree 1 map and $-1$ for the degree 5 map. The overall sign is the product, $-1$. For the cluster with $B=13$ there is a single relevant rational map of degree 13, for which the FR sign of a $\frac{\pi}{2}$ rotation is $-1$. For the $B=19$ octahedron, one may combine the rational maps of degrees $1,5$ and $13$. The two outer rational maps have FR signs $-1$, and the inner one $+1$, so the total sign is $+1$. 

\section{Spin/Isospin/Parity of Quantized States} 

\subsection{Quadrality 0 Clusters}

For quadrality 0 clusters, the baryon number is odd, so states must have half-integer spin and isospin. If there is only T symmetry, the constraints on quantum states can be written as
\bea
e^{i\pi L_3}e^{i\pi K_3} |\Psi \rangle &=& |\Psi
\rangle \,, \label{Tconstraint1} \\ 
e^{i\frac{2\pi}{3\sqrt{3}}(L_1+L_2+L_3)}
e^{i\frac{2\pi}{3\sqrt{3}}(K_1+K_2+K_3)} 
|\Psi \rangle &=& |\Psi \rangle \,. \label{Tconstraint2}
\eea
The first constraint imposes symmetry under a $\pi$ rotation, and the second under a $\frac{2\pi}{3}$ rotation. Together these symmetries generate the even part of T. As $M(R) = R$, the rotation and isorotation operators have the same form. The FR signs on the right hand side are all $+1$. The rotation and isorotation generators ${\bf L}$ and ${\bf K}$ (which mutually commute) can be combined into grand spin generators ${\bf M} = {\bf L} + {\bf K}$. The symmetry conditions (\ref{Tconstraint1}) and (\ref{Tconstraint2}) imply that $|\Psi \rangle$ is a singlet under the tetrahedral subgroup of SO(3) grand spin.
 
Recall that for a $B=1$ Skyrmion, the states are invariant under the full SO(3) of grand spin. This implies that the isospin $I$ and the spin $J$ have to be the same, and the state $|\Psi \rangle$ is the usual singlet of grand spin constructed by combining two equal angular momenta. Here slightly less is true. The grand spin can be $M=0$ but it can also be $M=3$ or $M=4$, as these SO(3) grand spin multiplets contain singlets under the tetrahedral subgroup. For $M=0$ the spin and isospin are the same, so the allowed combinations are $I =\frac{1}{2}$ with $J = \frac{1}{2}$, $I = \frac{3}{2}$ with $J =\frac{3}{2}$, and so on. For $M=3$, allowed combinations are $I =\frac{1}{2}$ with either $J = \frac{5}{2}$ or $J = \frac{7}{2}$, and $I = \frac{3}{2}$ with various $J$ from $\frac{3}{2}$ upwards, and also $I = \frac{5}{2}$ with various $J$ from $\frac{1}{2}$ upwards. For $M=4$ a new combination is $I = \frac{1}{2}$ with $J = \frac{9}{2}$, and there are many others.

In the Skyrme model, the parity operation is the combination of inversions in space and isospace. Clusters with only T symmetry do not have inversion symmetry, but the  Skyrmion has at least one reflection symmetry (a combination of a reflection in space and a reflection in isospace), so the parity operation can be re-expressed as a combination of rotations by $\pi$ in space and isospace, in the planes of these reflections. As there is a $B=1$ Skyrmion at the origin with orientation $1$, the relevant reflection planes are the same in space and isospace. Thus the parity operator is a combined $\pi$ rotation in space and isospace around an axis passing through opposite edge centres of a cube containing the tetrahedron. This is an element of the group O, outside its subgroup T. The parity eigenvalue therefore depends on the state. It is $+1$ for states with grand spin 0 or 4, which have the larger invariance under the group O, and $-1$ for states with grand spin 3. 

For clusters of quadrality 0 with symmetry O, the symmetry constraints (\ref{Tconstraint1}) and (\ref{Tconstraint2}) are replaced by
\bea
e^{i\frac{\pi}{2} L_3}e^{i\frac{\pi}{2} K_3} |\Psi \rangle &=& \pm |\Psi
\rangle \,, \label{Oconstraint1} \\ 
e^{i\frac{2\pi}{3\sqrt{3}}(L_1+L_2+L_3)}
e^{i\frac{2\pi}{3\sqrt{3}}(K_1+K_2+K_3)} 
|\Psi \rangle &=& |\Psi \rangle \,, \label{Oconstraint2}
\eea
where the right hand side of (\ref{Oconstraint1}) includes the FR sign for a $\frac{\pi}{2}$ rotation. Squaring the operator in (\ref{Oconstraint1}) reproduces the constraint (\ref{Tconstraint1}), so the tetrahedral subgroup constrains states in the same way as before. If the FR sign is $+1$, then the allowed states are those with grand spin 0 or 4, because for these there is an octahedral singlet. If the FR sign is $-1$ then the grand spin must be 3, because here the tetrahedral singlet transforms to its negative under a $\frac{\pi}{2}$ rotation. Clusters of quadrality 0 and symmetry O are invariant under a combined inversion, like the $B=1$  Skyrmion. The quantum states therefore all have parity eigenvalue $+1$. 

When the FR sign is $+1$, the allowed states for $I=\frac{1}{2}$ are therefore $J^P=\frac{1}{2}^+, \frac{7}{2}^+$ or $\frac{9}{2}^+$, for $I=\frac{3}{2}$ they are $J^P=\frac{3}{2}^+$ and above, for $I=\frac{5}{2}$ they are also $J^P=\frac{3}{2}^+$ and above, and for $I=\frac{7}{2}$ all half-integer spins are allowed. When the FR sign is $-1$, then the allowed states for $I=\frac{1}{2}$ are $J^P=\frac{5}{2}^+$ or  $\frac{7}{2}^+$, for $I=\frac{3}{2}$ they are $J=\frac{3}{2}^+$ and above, and for $I=\frac{5}{2}$ they are $J=\frac{1}{2}^+$ and above.
  
Examples with quadrality 0 and symmetry O, ignoring $B=1$, are the pure octahedra with $B=19$ and $B=85$ having FR sign $+1$, and the octahedra with $B=13, 55$ and $79$ having FR sign $-1$.

\subsection{Quadrality 2 Clusters}

The analysis of the quantum states of quadrality 2 clusters is rather similar. Both symmetry groups T and O can occur, and they act in the same way as for the quadrality 0 clusters. There are three cases, as before: T, O with all FR signs $+1$, and O with some FR signs $-1$. The allowed grand spins and the parity assignments are as before. The difference is that the baryon number $B$ is always even, so that the spin and isospin must be integers.

For clusters with only T symmetry, there are states with isospin 0 and spin/parity $J^P = 0^+, 3^-$ and $4^+$, corresponding to grand spins $0,3$ and $4$. For isospin 1, the lowest spin/parities are $J^P = 1^+, 2^-$ corresponding to grand spins 0 and 3. For isospin 2 there are more states, with the lowest being $J^P = 1^-, 2^+$ corresponding to grand spins 3 and 0. Finally for isospin 3 the lowest spin/parity state is $0^-$, with grand spin 3.

If the cluster has O symmetry and all FR signs $+1$, the allowed states are those with grand spins 0 or 4. In particular, the lowest spin/parities for isospins $0,1,2,3$ are respectively $J^P = 0^+, 1^+, 2^+, 1^+$. For a cluster with O symmetry and some FR signs $-1$, the grand spin must be 3, so the lowest allowed states for isospins $0,1,2,3$ are $J^P = 3^+, 2^+, 1^+, 0^+$. This last sequence should occur for the $B=38$ cluster, for example, and we will discuss this further below.  

\subsection{Quadrality 1 Clusters}

The remaining clusters are those with quadrality 1. These all have T symmetry, realised through the non-trivial 1-dimensional representation $R \to M(R)$, and the FR signs are all $+1$. The baryon number is always even, as it is a multiple of 4. The analysis of the allowed quantum states is not quite straightforward, but has been investigated before, in the context of the quantization of the $B=16$ Skyrmion and other Skyrmions with the same symmetry and baryon number a multiple of 4 \cite{Woo}. There are quite a lot of states with low isospin, and spins up to $J = 4$. We will not repeat the analysis. It was initially done using properties of Wigner functions, but was simplified in \cite{LM2} by considering the action of the group T on polynomials in the Cartesian coordinates $x,y,z$. The lowest spin/parity states for isospins $I = 0,1,2$ are $J^P = 0^+$, $0^-$, $0^+$, and for $I=3$ there are allowed $0^+$ and $0^-$ states. For isospin 0, the map $R \to M(R)$ plays no role, so the states are in the same, standard tetrahedral rotational band as for quadrality 2, with $J^P = 0^+, 3^-, 4^+, 6^\pm, 7^-,8^+,9^\pm, 10^\pm$, where we have included some of the higher spin states.

In summary, we have outlined the allowed states obtained by rigid body quantization for clusters with either T or O symmetry. There are seven cases. For both quadralities 0 and 2, there are three cases: T, ${\rm O}+$ and ${\rm O}-$, where the notation shows the symmetry group and the FR sign for a $\frac{\pi}{2}$ rotation. $B$ is odd for quadrality 0 and even for quadrality 2. For quadrality 1 the symmetry group is always T, and $B$ is even. 

\section{Energies of Quantized States}

Following the notation of \cite{MMW}, the kinetic energy of a quantum state is given by
\be
T = \frac{1}{2}\Bra{\Psi} H^T \mathcal{W}^{-1} H \Ket{\Psi} \, ,
\label{KineticEnergy}
\ee 
where $H^T=\left(K_1,K_2,K_3,L_1,L_2,L_3\right)$ is the vector of body-fixed isospin and spin operators, and
\be
\mathcal{W} = \begin{pmatrix} U & -W \\ -W^T & V \end{pmatrix}
\ee
is the overall moment of inertia tensor. The formula \eqref{KineticEnergy} is generally rather complicated but it simplifies considerably for the Skyrmions we consider. This is due to their symmetries which, as we have seen, lead to simple, diagonal inertia tensors.

For quadrality $0$ and $2$, the energy \eqref{KineticEnergy} of a state with spin $J$ and isospin $I$ becomes
\be
T= \frac{1}{2} \frac{1}{uv-w^2}\big( (u-w)J(J+1) +
  (v-w)I(I+1)+w M(M+1) \big)\,,
\label{Quad02energy}
\ee
where $u$, $v$ and $w$ were defined in Section 3, and $M$ is the grand spin. Note that $w$ is generally rather small compared to $u$ and $v$ and so the energy is close to that of uncoupled spherical tops. Quadrality $1$ Skyrmions are slightly more complicated as their isospin tensor has two independent diagonal entries, but $w=0$. We find that
\be
T = \frac{1}{2}\left( \frac{1}{v}J(J+1) + \frac{1}{U_{11}}I(I+1) 
+\left( \frac{1}{U_{33}} - \frac{1}{U_{11}}\right)K_3^2\right) \,,
\label{Quad1E}
\ee
where $K_3$ is the (eigenvalue of the) third component of the  body-fixed isospin. Recall that $K_3$ is not directly observable, unlike the ``space-fixed'' $I_3$, which, together with $B$, determines the proton and neutron numbers.
 
For large $B$ the moments of inertia grow like
\be
u \sim B \ \text{and} \ v \sim B^{5/3} \, ,
\ee
providing a separation of scales. The isospin contribution is much larger than the spin contribution. Hence some concepts often used in the Skyrme model, such as rotational bands, are not as simple here. For small $B$, energy levels are usually first ordered by their isospin $I$, then by $J$, with the states lying on an approximate rotational band proportional to $J(J+1)$. With this information, one immediately knows that a larger spin leads to a larger energy. Also, the ratios between the energies of the states are simply related. This picture does not hold for large $B$. As an extreme example, consider the $B=80_{1b}$ Skyrmion. For $I=3$ this has a spin $3$ state with $K_3=0$. Its energy in Skyrme units, calculated using equation \eqref{Quad1E} and values from Table \ref{NumericalTable}, is
\be
T_{J=3, I=3} = \frac{1}{2}\left(\frac{12}{139000} + \frac{12}{2900} \right) =  2.11 \times 10^{-3} \, .
\ee
There is also a spin $4$ state with $K_3=3$ that has energy 
\be
T_{J=4, I=3} = \frac{1}{2}\left(\frac{20}{139000} + \frac{12}{2900} + \left( \frac{1}{3180} - \frac{1}{2900}\right)9 \right) =  2.00 \times 10^{-3} \, .
\ee
Hence, unusually, the state with higher spin has lower energy. Large nuclei sometimes have several rotational bands in their experimental energy spectrum, or do not have easily distinguishable rotational bands. This analysis demonstrates one reason why. However, for small $I$, the concept of simple rotational bands remains valid as we will demonstrate below.

To compare our results to experimental data we must convert equations \eqref{Quad02energy} and \eqref{Quad1E} from Skyrme units to physical ones. To do this, we use the asymmetry term of the semi-empirical mass formula. In our notation, this term is
\be
a_A \frac{(2I)^2}{B}
\label{semiemp-asym}
\ee
where $a_A$ has a physical value of around $23.2$ MeV. The isospin energy contribution in the Skyrme model is approximately
\be
\frac{I(I+1)}{2u} \, .
\ee 
Since the isospin inertia $u$ scales as $B$, this matches the asymmetry term \eqref{semiemp-asym} for large nuclei. By comparing these formulae for the $B=80$ Skyrmions we find an energy conversion factor of
\be
7300 \, \text{MeV} \, .
\ee
Hence the states discussed in the previous paragraph have kinetic energy contributions of $15.40$ MeV and $14.60$ MeV respectively. This is a new method to calibrate the Skyrme model which we hope will give good results across a large range of nuclei.

Let us now turn to some examples, and compare the Skyrmion predictions with experimental data. By restricting to rigid body quantization we have ignored some physics which will be important in describing the full energy spectrum. Many more states would arise if we had quantised the Skyrmion's vibrational modes. Some of these modes will be soft, especially since Coulomb energy, which we have ignored, favours configurations which are ellipsoidal rather than those we have considered. Hence, we do not expect to reproduce the full experimental energy spectrum for the nuclei, but hope to identify some of the low energy states with those we have calculated. 

The $B=31$ Skyrmion is a truncated tetrahedron of quadrality $0$, and is relatively tightly bound. Experimentally, nuclei with different isospins $I$ are distinct, and for given $I$ the nucleus that is most neutron-rich has $I_3 =-I$. The neutron-rich sequence for $B=31$ is $^{31}{\rm P}$, $^{31}{\rm Si}$, $^{31}{\rm Al}$, and $^{31}{\rm Mg}$, with isospins $I = \frac{1}{2}, \frac{3}{2}, \frac{5}{2}$ and $\frac{7}{2}$. As explained in Section 5, the Skyrmion quantum states have grand spins $0, 3, 4, \dots$. The energy of a state with spin $J$, isospin $I$ and grand spin $M$ is given by equation \eqref{Quad02energy}. Owing to the sizes of the moments of inertia, the states are energetically ordered first by their isospin, then by spin and finally by grand spin. Since the $B=31$ Skyrmion has $w = -113 < 0$, a large $M$ is favoured. This fact provides a mechanism for high spin states (with large $M$) to have less energy than low spin states (with small $M$). However, $w$ is too small for this mechanism to change any orderings for the low energy states of this particular Skyrmion. We tabulate the lowest energy states in Table \ref{B31results} and compare them to the corresponding nuclei. The energies are measured relative to the static mass of the Skyrmion. Our results for the ground state spins of the Phosphorus-31 and Magnesium-31 nuclei match the experimental data. The Magnesium-31 nucleus is difficult to probe experimentally. The ground state spin/parity assignment of $\frac{1}{2}^+$ has only recently been clarified \cite{NKY}. For a history of the experimental and theoretical work on the nucleus see \cite{Ney}. The nucleus is a candidate member of the ``island of inversion" due to the unusual ground state. This state is not predicted by the traditional shell model. Instead it is an ``intruder state", which can only be described once the shell model interaction is modified \cite{Mar}. For us, the spin/parity $\frac{1}{2}^+$ assignment of the ground state is a simple consequence of the symmetry of the nucleus. We also find two spin $\frac{3}{2}$ states with a small splitting, which are seen in the data. However, we find a low energy $\frac{1}{2}^-$ state which is not seen. Our model of Silicon-31 has a ground state spin/parity $\frac{3}{2}^-$, inconsistent with the experimental ground state which has spin/parity $\frac{3}{2}^+$. Instead, the true ground state is described by our first excited state, which lies only $0.16$ MeV above the model ground state. There is a $\frac{5}{2}^+$ state in the experimental data whose transition rate to the $\frac{3}{2}^+$ ground state is large, around $12$ Weisskopf units. Such a value suggests that the states are related and have a collective nature - matching our interpretation of the states as rotational excitations of the Skyrmion.

\bgroup
\def \arraystretch{1.2}
\begin{table}
	\begin{center}
		\begin{tabular}{|c|c|c|c|c|c|c|}
			\hline
			$I$ & $J^P$ & $M$ & $T$ (MeV) & Nucleus & Ground state Spin/Parity & Match? \\  \hline
			$\frac{1}{2}$ & $\frac{1}{2}^+$ & 0 & 2.36 & $^{31}$P & $\frac{1}{2}^+$ & Y \\ 
			& $\frac{5}{2}^-$ & 3 & 3.47 & & &  \\
			& $\frac{7}{2}^+$ & 4 & 4.48 & & &  \\
			& $\frac{7}{2}^-$ & 3 & 4.59 & & &  \\ \hline
			$\frac{3}{2}$ & $\frac{3}{2}^-$ & 3 & 11.62 & $^{31}$Si & $\frac{3}{2}^+$ & N \\ 
			& $\frac{3}{2}^+$ & 0 & 11.78 & & &  \\ 
			& $\frac{5}{2}^+$ & 4 & 12.31 & & & \\
			& $\frac{5}{2}^-$ & 3 & 12.42 & & &   \\ \hline
			$\frac{5}{2}$ & $\frac{1}{2}^-$ & 3 & 26.05 & $^{31}$Al & $\frac{5}{2}^+$ & N \\ 
			& $\frac{3}{2}^+$ & 4 & 26.42 & & &  \\ 
			& $\frac{3}{2}^-$ & 3 & 26.53 & & &  \\
			& $\frac{5}{2}^+$ & 4 & 27.22 & & &  \\ \hline
			$\frac{7}{2}$ & $\frac{1}{2}^+$ & 4 & 46.82 & $^{31}$Mg & $\frac{1}{2}^+$ & Y \\ 
			& $\frac{1}{2}^-$ & 3 & 46.93 & & &  \\ 
			& $\frac{3}{2}^+$ & 4 & 47.30 & & &  \\
			& $\frac{3}{2}^-$ & 3 & 47.41 & & &  \\ \hline
		\end{tabular}
		\vskip 7pt
		\caption{Low energy states of the $B=31$ Skyrmion.}
		\label{B31results}
	\end{center}
\end{table}
\egroup

The $B=38$ Skyrmion has quadrality $2$ and symmetry ${\rm O}-$. Its quantum states also have energy given by \eqref{Quad02energy}. Again $w<0$, so a large grand spin is energetically favoured. Due to the octahedral symmetry of the Skyrmion, states with grand spin $0$ are ruled out. Instead, they must have grand spin $3, 6, \dots$ . The low energy quantum states are shown in Table \ref{B38results}. Here, two ground states have the correct spin. Most notably, we find that the ground state spin for Potassium-38 is $3^+$, agreeing with the experimental result. Our model wrongly predicts the ground state of Argon-38 to have spin $2^+$. However, the states of Argon-38 are curious. Just above the $0^+$ ground state there are several rotational bands, with spins 2 and upwards. Due to the symmetry of the $B=38$ Skyrmion, all of the predicted states have positive parity. This is a major shortcoming of our calculation since the $B=38$ nuclei have many experimental states with negative parity. In particular, the Chlorine-38 ground state has spin/parity $2^-$. We could include negative parity states by allowing the Skyrmion to deform and including its vibrational modes in the quantization. For instance, there is a vibrational mode which breaks octahedral but retains tetrahedral symmetry. Coupling this mode to the rotations will allow for negative parity states.

\bgroup
\def \arraystretch{1.2}
\begin{table}
	\begin{center}
		\begin{tabular}{|c|c|c|c|c|c|c|}
			\hline
			$I$ & $J$ & $M$ & $T$ (MeV) & Nucleus &
                        Ground state Spin/Parity & Match? \\  \hline
			$0$ & $3^+$ & 3 & 1.17 & $^{38}$K & $3^+$ & Y \\ 
			& $6^+$ & 6 & 4.09 & & &  \\ 
			& $7^+$ & 7 & 5.45 & & &  \\
			& $9^+$ & 9 & 8.76 & & &  \\ \hline
			$1$ & $2^+$ & 3 & 5.49 & $^{38}$Ar & $0^+$ & N \\ 
			& $3^+$ & 3 & 6.12 & & &  \\ 
			& $4^+$ & 3 & 6.96 & & &  \\
			& $5^+$ & 6 & 7.78 & & &  \\ \hline
			$2$ & $1^+$ & 3 & 14.97 & $^{38}$Cl & $2^-$ & N \\ 
			& $2^+$ & 3 & 15.39 & & &  \\ 
			& $3^+$ & 3 & 16.01 & & & \\
			& $4^+$ & 6 & 16.63 & & &  \\ \hline
			$3$ & $0^+$ & 3 & 29.60 & $^{38}$S & $0^+$ & Y \\ 
			& $1^+$ & 3 & 29.81 & & &  \\ 
			& $2^+$ & 3 & 30.23 & & &  \\
			& $3^+$ & 6 & 30.63 & & &  \\ \hline
		\end{tabular}
		\vskip 7pt
		\caption{Low energy states of the $B=38$ Skyrmion.}
		\label{B38results}
	\end{center}
\end{table}
\egroup

Finally, consider the quadrality 1, $B=80_{1b}$ Skyrmion in an $I=0$ state. This is a model for Zirconium-80. Rigid body quantization allows for states with spin/parity $0^+, 3^-, 4^+, 6^\pm, \dots$. Their energies are simply
\be
\frac{J(J+1)}{2v} \, 7300 \text{ MeV} \, .
\ee
This gives a pure, tetrahedral rotational band with energies $0, 0.32, 0.53, 1.10, \dots$ MeV. Experimentally, Zirconium-80 does have a rotational band, though with spins $0^+, 2^+, 4^+,$ $6^+, 8^+, 10^+$ and energies $0, 0.29, 0.83, 1.61, 2.61, 3.79$ MeV. These are not the states described by our tetrahedral Skyrmion. To include these states in our model we must consider a different low energy $B=80$ Skyrmion or allow the tetrahedron to deform. However, the predicted energies are of the same order as the experimental states. This gives us confidence in the new calibration that we have suggested.  The energies from the experimental rotational band suggest that the intrinsic shape of the nucleus is highly deformed, but it is possible that further states of this hard-to-produce, semi-magic $Z=N$ nucleus with less deformation are still to be discovered. Other authors have suggested that a tetrahedral intrinsic structure is favoured \cite{TYM,DGSM,TSD}, and this is also the prediction of the Skyrme model.

\section{Conclusions}

We have used FCC cluster Skyrmions of the lightly-bound Skyrme model as starting configurations in our search for solutions of the standard Skyrme model. This paper has focussed on those clusters that correspond to SU(4) weight diagrams, which are all tetrahedrally symmetric subclusters of the FCC lattice, and sometimes octahedrally symmetric. The Skyrmions we have found are mostly novel, and have a range of even and odd baryon numbers up to $B=85$. SU(4) representation theory allows us to find algebraic formulae that approximately describe some physical properties of the classical Skyrmions, including their binding energies.

We have discussed the rigid body quantization of these Skyrmions. Because of the symmetries, just a few patterns of allowed spin/isospin/parity states occur. The interest of our treatment, compared to standard treatments of collective motion in nuclear physics, is that spin and isospin occur in a unified manner as coupled collective excitations. This is not novel in the context of Skyrmions, but occurs here in some novel ways, and is applied to several new examples of Skyrmions with relatively large baryon number. We have also presented the formulae for the energies of these states; they depend on a few moments of inertia that we have calculated numerically. We have discussed in some detail a few examples of the Skyrmion states that arise, in particular for $B=31$, $B=38$ and $B=80$, and compared the results with experimentally known states.

Further work is needed, to investigate more examples of Skyrmions arising from weight diagrams, but also to consider competing clusters of low energy with different structures. Intermediate baryon numbers should also be considered. They arise by attaching or removing one or more $B=1$ Skyrmions near the surface of a symmetric cluster.

\vspace{7mm}

\section*{Acknowledgements}

We are grateful to Amit Hazi for informing us of reference \cite{Kas}. We are also grateful to Chris Lau for constructing the $B=40$ Skyrmion using a three-layer rational map ansatz, obtaining a result very similar to that in Figure \ref{Q1sk}. NSM thanks David Jenkins and his colleagues at York for discussions.
 
This work has been partially supported by STFC consolidated grant ST/P000681/1. CJH is supported by the National Natural Science Foundation of China grant 11675223. JIR is supported by an EPSRC studentship.

\end{document}